\documentclass{plateau}

\usepackage{xcolor}

\title{Literate Execution}

\plateauConference[PLATEAU '26]{16}{March 9--10, 2025}{Pittsburgh, PA, USA}
\setcopyright{none}
\acmDOI{}
\usepackage{float}
\usepackage{mathpartir}
\usepackage{bm}
\usepackage{soul}
\usepackage{xspace}
\usepackage{mathtools}
\usepackage{tikz}
\usepackage{tikz-cd}
\usetikzlibrary{positioning, shapes, arrows, fit}
\usepackage[scaled=0.75]{beramono}
\usepackage{calrsfs}
\usepackage{subdepth}
\usepackage{dashundergaps}
\usepackage[final]{listings}
\usepackage{mleftright}
\usepackage{extarrows}
\usepackage{placeins}
\usepackage[a]{esvect}
\usepackage{colortbl}
\usepackage{tabularx}
\usepackage{multirow}
\usepackage{subcaption}
\usepackage{enumitem}
\usepackage{longtable}
\usepackage{pifont}
\usepackage{cancel}
\usepackage{wrapfig}
\usepackage[color]{changebar}
\usepackage{marginnote}

\newcolumntype{L}[1]{>{\raggedright\let\newline\\\arraybackslash\hspace{0pt}}p{#1}}
\newcolumntype{C}[1]{>{\centering\arraybackslash}p{#1}}

\newcommand*{\ala}{\emph{\`a la}\xspace}

\newcommand*{\ttt}[1]{\texttt{#1}}
\newcommand*{\kw}[1]{{\text{\ttt{#1}}}} 

\newcommand*{\twoPrime}{{\prime\mkern-2.6mu\prime\mkern-2.2mu}}

\DeclareTextFontCommand{\textbfit}{%
  \fontseries\bfdefault 
  \itshape
}

\newcommand*{\concat}{\cdot}

\newcommand*{\dom}[1]{\textsf{dom}(#1)}

\newcommand*{\eqdef}{\stackrel{\smash{\text{\tiny def}}}{=}}

\newcommand*{\set}[1]{\{#1\}}

\newcommand*{\figref}[1]{Figure~\ref{fig:#1}}

\newcommand*{\Secref}[1]{\S\,\ref{sec:#1}}

\newenvironment{nop}{}{}
\newenvironment{sdisplaymath}
   {\begin{nop}\small\begin{displaymath}}
   {\end{displaymath}\end{nop}\ignorespacesafterend}
\newenvironment{smathpar}
   {\begin{nop}\small\begin{mathpar}}
   {\end{mathpar}\end{nop}\ignorespacesafterend}

\newenvironment{mathfig}{\begin{sdisplaymath}}{\end{sdisplaymath}}

\makeatletter
\newbox\sf@box
  {\def\sf@one{#1}%
   \def\sf@two{#2}%
   \setbox\sf@box\hbox
      \bgroup}%
  { \egroup
   \ifx\@empty\sf@two\@empty\relax
     \def\sf@two{\@empty}
   \fi
   \ifx\@empty\sf@one\@empty\relax
      \subfloat[\sf@two]{\box\sf@box}%
   \else
      \subfloat[\sf@one][\sf@two]{\box\sf@box}%
   \fi}
\makeatother

\RequirePackage{xcolor}

\definecolor{highlightcolor}{rgb}{1.0,0.8,0.8}
\definecolor{shadecolor}{rgb}{0.9,0.9,0.9}
\definecolor{lightgray}{rgb}{0.8,0.8,0.8}
\newcommand*{\shadebox}[1]{\fcolorbox{lightgray}{shadecolor}{\raisebox{0pt}[0.60\baselineskip][0.05\baselineskip]{#1}}}

\newcommand*{\ruleName}[1]{\textnormal{\textsf{#1}}}

\makeatletter
\newcommand{\superimpose}[2]{%
  {\ooalign{$#1\@firstoftwo#2$\cr\hfil$#1\@secondoftwo#2$\hfil\cr}}}
\makeatother

\newsavebox{\vardisplaymathbox}

\definecolor{addedcolor}{RGB}{0,160,0}
\definecolor{deletedcolor}{RGB}{180,180,180}

\makeatletter
\newcommand{\cboff}{\let\end@dblfloat\ltx@end@dblfloat}
\newcommand{\cbon}{\let\end@dblfloat\cb@end@dblfloat}
\makeatother

\newcommand*{\OurLang}{Fluid\xspace} 

\newcommand{\bind}[2]{{#1}:{#2}}
\newcommand*{\seq}[1]{\vv{#1}}
\newcommand*{\nonEmpty}[1]{#1^{\raisebox{0.2ex}{\scalebox{0.6}{$+$}}}}
\newcommand*{\seqEmpty}{\varepsilon}
\newcommand*{\mathSf}[1]{\textup{\textsf{#1}}} 
\newcommand*{\blockSeq}[1]{\overline{#1}}
\newcommand*{\sym}[1]{\textcolor{blue}{\kw{#1}}}
\newcommand*{\seqRange}[2]{\seqRangeOp{#1}{#2}{,\,}}
\newcommand*{\seqRangeOp}[3]{{#1} #3 \iter #3 {#2}}
\newcommand*{\iter}{..}
\newcommand*{\datatype}[1]{\Sigma(#1)}
\newcommand*{\arity}[1]{\mathSf{arity}(#1)}
\newcommand*{\envEmpty}{\varnothing}

\newcommand*{\comma}{\sym{,}}
\newcommand*{\equal}{\mathrel{\sym{=}}}
\newcommand*{\dblQuote}{\sym{\textquotedbl}}

\newcommand*{\exApp}[2]{{#1}\sym{(}{#2}\sym{)}}
\newcommand*{\exBinaryApp}[3]{{#1} \mathbin{#2} {#3}}
\newcommand*{\exConstr}[2]{#1\sym{(}#2\sym{)}}
\newcommand*{\exDef}[2]{\sym{def}\;{p}\sym{:}\;{#2}}

\newcommand*{\exDict}[1]{\sym{\{}{#1}\sym{\}}}
\newcommand*{\exDoc}[2]{\sym{@doc}\sym{(}{#1}\sym{)}\;{#2}}
\newcommand*{\exDocCore}[2]{\sym{doc}\sym{(}{#1},{#2}\sym{)}}
\newcommand*{\exDProject}[2]{{#1}\sym{[}{#2}\sym{]}}
\newcommand*{\exFirstClassOp}[1]{\sym{(}{#1}\sym{)}}
\newcommand*{\exFun}[2]{\sym{lambda}\;{#1}\sym{:}\;{#2}}
\newcommand*{\exLambda}[1]{\lambda{#1}}
\newcommand*{\exIfThenElse}[3]{\sym{if}\;{#1}\sym{:}\;{#2}\;\sym{else:}\;{#3}}
\newcommand*{\exInfixFun}[1]{{\sym{|}}#1{\sym{|}}}
\newcommand*{\exList}[1]{\sym{[}{#1}\sym{]}}
\newcommand*{\exListComp}[2]{\sym{[}\,{#1}\;{#2}\sym{]}}
\newcommand*{\exListEnd}{\kw{]}}
\newcommand*{\exListNext}[2]{\comma\;#1 \ #2}
\newcommand*{\exMatch}[2]{\sym{match}\;{#1}\sym{:}\;{#2}}
\newcommand*{\exParagraph}[1]{\sym{f}\dblQuote\dblQuote\dblQuote{#1}\dblQuote\dblQuote\dblQuote}
\newcommand*{\exProject}[2]{{#1}\sym{.}{#2}}
\newcommand*{\exClosure}[3]{\kw{cl}({#1},{#2},{#3})}
\newcommand*{\exImport}[1]{\sym{import}\;#1}

\newcommand*{\exLet}[3]{\kw{let}\;{#1}\equal{#2}\;\kw{in}\;{#3}}
\newcommand*{\exLetTL}[2]{\kw{let}\;{#1}\equal{#2}}
\newcommand*{\exLetRec}[2]{\kw{let}\;{#1}\;\kw{in}\;{#2}}
\newcommand*{\paraToken}[1]{#1}
\newcommand*{\paraUnquote}[1]{\sym{\{}{#1}\sym{\}}}

\newcommand*{\clauseFun}[3]{\sym{def}\;{#1}\sym{(}{#2}\sym{):}\;{#3}}
\newcommand*{\clauseMatch}[2]{\sym{case}\;{#1}\sym{:}\;{#2}}
\newcommand*{\clause}[2]{{#1}\equal{#2}}
\newcommand*{\clauseWith}[1]{\langle\mathbin{#1}\rangle}
\newcommand*{\append}{++}
\newcommand*{\clauseCons}{\clauseWith{\cons}}
\newcommand*{\clauseConcat}{\clauseWith{\append}}
\newcommand*{\pattConstr}[2]{\exConstr{#1}{#2}}
\newcommand*{\pattList}[1]{\exList{#1}}
\newcommand*{\pattRecord}[1]{\exDict{#1}}
\newcommand*{\pattVar}[1]{#1}
\newcommand*{\pattListEnd}{\exListEnd}
\newcommand*{\pattListNext}[2]{\exListNext{#1}{#2}}

\newcommand*{\qualDeclaration}[2]{\sym{def}\;{#1}\sym{:}\;{#2}}
\newcommand*{\qualGenerator}[2]{\sym{for}\;{#1}\;\sym{in}\;{#2}}
\newcommand*{\qualGuard}[1]{\sym{if}\;#1}

\newcommand{\opName}[1]{\mathsf{#1}}

\newcommand*{\inStar}[2]{\set{#1 \mapsto #2}}

\newcommand*{\fresh}[2]{#1 \notin \V(#2)}
\newcommand*{\V}{\opName{V}}

\newcommand*{\evalR}{\Rightarrow}
\newcommand*{\decEval}[1]{\overset{\tiny\lower.5em\hbox{#1}}{\evalR}}
\newcommand*{\evalM}{\decEval{m}}
\newcommand*{\evalD}{\decEval{d}}
\newcommand*{\closeDefs}{\rightarrowtail}
\newcommand*{\interpret}[1]{\hat{#1}}
\newcommand*{\match}{\rightsquigarrow}
\newcommand*{\orElse}{\rightharpoonup}

\newcommand*{\desugar}{\twoheadrightarrow}

\newcommand{\cons}{\cdot}
\newcommand*{\length}[1]{|#1|}
\newcommand*{\numleq}{\leq}

\newcommand*{\annot}[2]{#1_{#2}}

\newcommand*{\annClosure}[4]{\annot{\exClosure{#1}{#2}{#3}}{#4}}

\newcommand*{\annConstr}[3]{\annot{\exConstr{#1}{#2}}{#3}}

\newcommand*{\annDict}[2]{\annot{\exDict{#1}}{#2}}

\newcommand*{\annInt}[2]{\annot{#1}{#2}}

\newcommand*{\elimmapsto}{\mapsto}
\newcommand*{\elimBind}[2]{{#1}\elimmapsto{#2}}
\newcommand*{\elimConstr}[1]{\set{#1}}
\newcommand*{\elimDict}[2]{\elimBind{\exDict{#1}}{#2}}
\newcommand*{\elimVar}[2]{\elimBind{#1}{#2}}

\newcommand*{\cCons}{\kw{Cons}}
\newcommand*{\cNil}{\kw{[]}}
\newcommand*{\cFalse}{\kw{False}}
\newcommand*{\cTrue}{\kw{True}}
\newcommand*{\cPara}{\kw{Paragraph}}

\newcommand*{\varAnon}{\kw{\_}}
\newcommand*{\varConcatMap}{\kw{concatMap}}

\newcommand*{\ctrsF}{\opName{ctrs}}

\definecolor{verylightgray}{gray}{0.9}
\definecolor{lightgray}{gray}{0.5}
\definecolor{mediumgray}{gray}{0.45}

\newlength\lsthorizontalpadding
\newcommand*\lstnumberstyle{\ttfamily\scriptsize\textcolor{lightgray}}
\newcommand*{\lbbar}{\{\kern-0.3em|}
\newcommand*{\rbbar}{|\kern-0.3em\}}
\newlength\lstnumbersep
\newlength\lstnumberwidth
\lstdefinelanguage{Fluid}{%
   morekeywords={as,else,fun,if,in,let,match,then}%
  ,moredelim=[s][\itshape]{`}{`}
}
\lstnewenvironment{codecol}[1][]{\lstset{language=code,#1,moredelim=*[is][\textcolor{mediumgray}]{|}{|}}}{}

\definecolor{base00}{HTML}{ffffff} 
\definecolor{base01}{HTML}{e0e0e0} 
\definecolor{base02}{HTML}{d6d6d6} 
\definecolor{base03}{HTML}{8e908c} 
\definecolor{base04}{HTML}{969896} 
\definecolor{base05}{HTML}{4d4d4c} 
\definecolor{base06}{HTML}{282a2e} 
\definecolor{base07}{HTML}{1d1f21} 
\definecolor{base08}{HTML}{c82829} 
\definecolor{base09}{HTML}{f5871f} 
\definecolor{base0A}{HTML}{eab700} 
\definecolor{base0B}{HTML}{718c00} 
\definecolor{base0C}{HTML}{3e999f} 
\definecolor{base0D}{HTML}{4271ae} 
\definecolor{base0E}{HTML}{8959a8} 
\definecolor{base0F}{HTML}{a3685a} 

\lstdefinestyle{base16python}{
  language=Python,
  backgroundcolor=\color{base00},
  basicstyle={\ttfamily\small\color{base05}},
  keywordstyle=\color{base0E},
  commentstyle=\color{base03},
  stringstyle=\color{base0B},
  numberstyle=\scriptsize\color{base03},
  identifierstyle=\color{base05},
  showstringspaces=false,
  breaklines=true,
  postbreak=\space,
  frame=none,
  rulecolor=\color{base01},
  tabsize=4,
  columns=fullflexible,
  keepspaces=true,
  moredelim=[l][\color{base0B}]{@}
}

\lstnewenvironment{python}[1][]{\lstset{style=base16python,#1}}{}
\newcommand{\pyinline}[1]{\lstinline[style=base16python]!#1!}

\begin{document}

\begin{abstract}
\emph{Literate programming}, introduced by \citet{knuth1984}, interleaves code and prose so that a program
can be read as both executable and explanatory text. We propose \emph{literate execution}, which inverts this
relationship: rather than embedding code within a static narrative, we treat documentation --- and other
expository elements such as visualisations --- as first-class artefacts that can be computed alongside a
running program and then integrated into a view of its execution. We explore this idea through Fluid, a
programming language with a provenance-tracking runtime that records fine-grained dependencies between inputs
and outputs. These provenance relationships can be surfaced as interactions that allow readers to explore how
intermediate values contribute to a result. By integrating visualisation, provenance, and exposition, literate
execution aims to make programs more explorable and self-explanatory, and explorable explanations easier to
program.
\end{abstract}

\author{Joe Bond}
\orcid{0009-0008-5995-6994}
\affiliation{%
  \institution{University of Bristol}
  \city{Bristol}
  \country{UK}
}

\author{Jacob Pake}
\orcid{0009-0001-3568-5383}
\affiliation{%
  \institution{University of Kent}
  \city{Canterbury}
  \country{UK}
}

\author{Cristina David}
\orcid{0000-0002-9106-934X}
\affiliation{%
  \institution{University of Bristol}
  \city{Bristol}
  \country{UK}
}

\author{Andrew McNutt}
\orcid{0000-0001-8255-4258}
\affiliation{%
  \institution{University of Utah}
  \city{Salt Lake City}
  \country{USA}
}

\author{Trevor Sseguya Muwonge}
\orcid{0009-0007-3294-9172}
\affiliation{%
  \institution{University of Bristol}
  \city{Bristol}
  \country{UK}
}

\author{Dominic Orchard}
\orcid{0000-0002-7058-7842}
\affiliation{%
  \institution{University of Cambridge}
  \city{Cambridge}
  \country{UK}
}
\affiliation{%
  \institution{University of Kent}
  \city{Canterbury}
  \country{UK}
}

\author{Roly Perera}
\orcid{0000-0001-9249-9862}
\affiliation{%
  \institution{University of Cambridge}
  \city{Cambridge}
  \country{UK}
}
\affiliation{%
  \institution{University of Bristol}
  \city{Bristol}
  \country{UK}
}

\maketitle

\section{From Literate Programming to Literate Execution}
\label{sec:introduction}

Data-driven documents, such as scientific papers and policy reports, typically present their findings
without exposing how those findings were produced. A reader who encounters a statistic or a chart has no way
to interrogate what went into it, or how it relates to the underlying data. The document is \emph{opaque}: one
sees the output of a computation, but not the computation itself.

There are various ways one might approach the challenge of making documents like these more
transparent. \emph{Explorable explanations}, popularised by \citet{victor2011a} and venues such as
\url{distill.pub} and \href{https://visxai.io/}{VISxAI}~\cite{hohman2020}, invite readers to adjust parameters
and see how results change, exploring how the computation would behave under different inputs or assumptions.
We call this \emph{counterfactual} transparency; multiverse analyses~\cite{steegen2016,dragicevic2019} pursue
a similar idea in the context of statistical reporting. But consider a report such as the IPCC's \emph{Summary
for Policymakers}~\cite{ipcc2021}. A policymaker reading it might want to understand not only how results
would differ under different assumptions, but also how the reported conclusions for a \emph{given} set of
assumptions follow from the data: what were the intermediate aggregation steps, and which parts of the input
contributed to a particular summary statistic? We call these sorts of concerns \emph{decompositional}
transparency. It has two axes: \emph{sequential} decomposition reveals the intermediate artefacts that arose
during the computation, while \emph{parallel} decomposition reveals which parts of the input influenced which
parts of the output. In this paper we argue that decompositional transparency is also an important facet of
transparency, and one largely unaddressed by existing tools.

Creating transparent documents of any kind is genuinely hard. Some of the difficulty is what
\citet{brooks1987} would call \emph{essential}: choosing what to explain, how to structure a narrative, and
which interactions will be pedagogically effective requires communicative skill that cannot be engineered
away. But alongside this essential difficulty, there is a large component of \emph{accidental} complexity.
Authoring systems such as Idyll~\cite{conlen2018} and Living Papers~\cite{heer2023} have provided valuable
infrastructure for supporting counterfactual transparency, reducing the effort needed to create explorable,
parameter-driven documents. Yet making a computation's internal structure transparent remains largely a manual
undertaking: the author must build bespoke interactive presentations that reimagine, and to some extent
reimplement, aspects of the computation for the purpose of explanation. This accidental complexity is closely
related to a familiar problem in software engineering: understanding why a program produced a given output, by
inspecting intermediate states and tracing dependencies, is the essence of debugging. The tools and techniques
that support decompositional transparency for end-users and those that support debugging for developers
address fundamentally the same problem.

In this paper we propose \emph{literate execution}, an approach that provides infrastructure for decompositional transparency. Rather than asking authors to build bespoke interactive presentations, we embed documentation directly into the execution of a program, so that the program's own dependency structure can generate interactions automatically. The name reflects a structural parallel with \emph{literate programming}~\cite{knuth1984}, which interleaves prose with static source code to make programs readable. Literate execution applies the same principle to a dynamic rather than static object: it interleaves prose with the execution, making the computation, not just the program, a literate artefact.


\begin{figure}[t]
  \includegraphics[width=\linewidth]{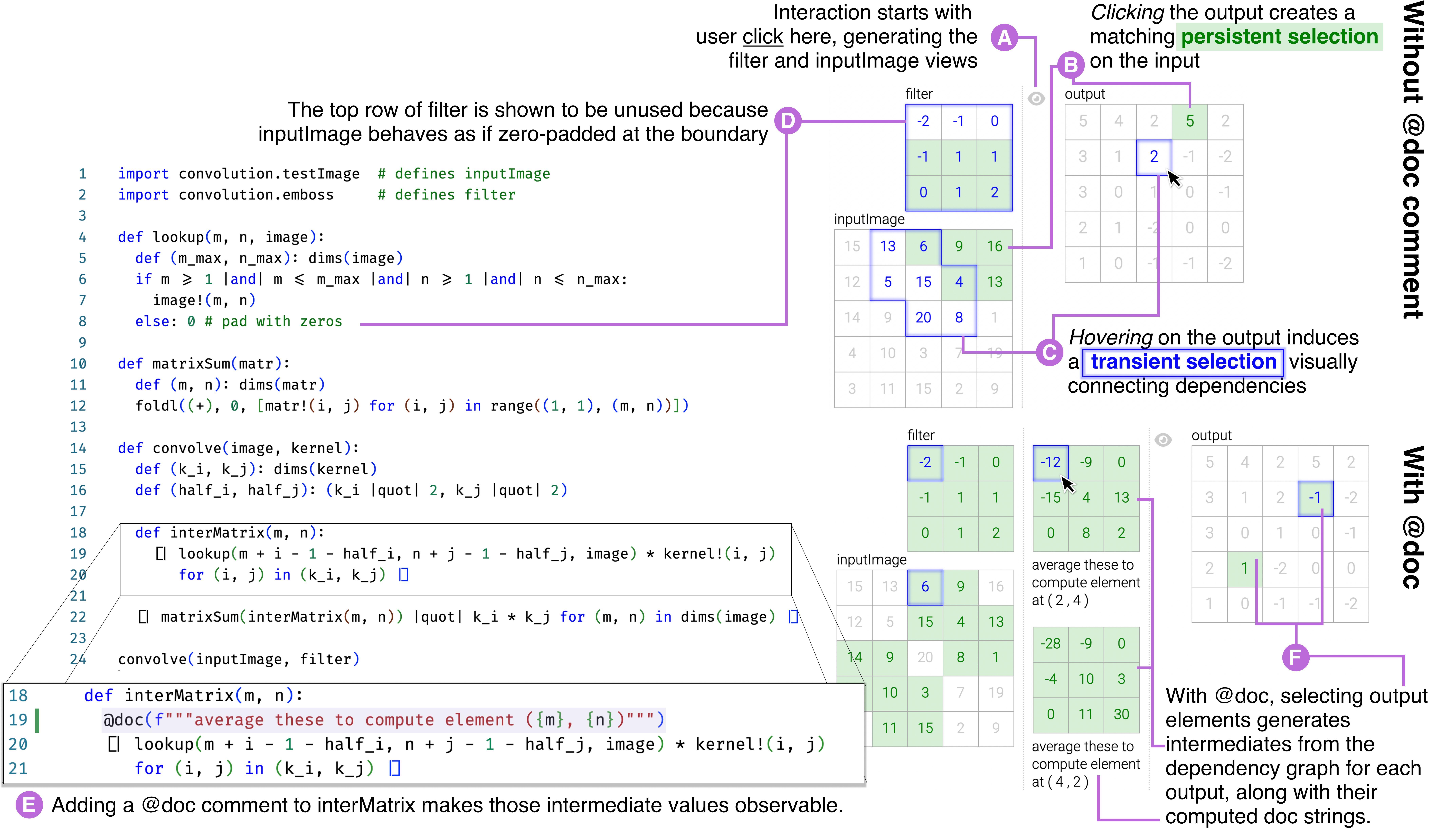}
    \vspace{-1em}
  \caption{This work extends our prior work~\cite{perera2022, bond2025} on automatic visualisation of dependencies by introducing `doc comments' and intermediate value exploration to facilitate \emph{literate execution}. Here we show how these new features can be used to explain matrix convolution.}
  \label{fig:convolution-intermediates}
  \vspace{-1em}
\end{figure}

We explore this idea using a programming language called Fluid (\url{http://f.luid.org}) which has built-in
dependency tracking~\cite{perera2022,bond2025}. In Fluid, every computation produces a \emph{dynamic dependence
graph} (DDG) alongside the computed output, capturing fine-grained dependencies between outputs and inputs. This
provenance information can then be made available to end-users via additional interactions, making it possible to
point at computed content and see, \emph{in situ} (for example on a web page hosting visualisations or text
authored in Fluid), how parts of the input contributed to various parts of the output. \figref{convolution-intermediates}
shows how a purely functional implementation of \emph{matrix convolution} in Fluid, an algorithm used in
machine learning and image processing, can be turned into an interactive figure automatically.
%
This paper introduces two new Fluid features for authoring transparent computational documents. First,
dynamically computed \textbf{doc comments}, possibly containing embedded visualisations, can be attached to
runtime values using a special decorator notation. Second, a UI for \textbf{exploring intermediate
values} that allows a user to interactively inspect them and trace their upstream and
downstream dependencies. We call the combination of these two features into a document authoring tool
\emph{literate execution}. Alongside these features we also present a new, Python-like surface syntax for Fluid, reflecting our target audience of data scientists and domain experts accustomed to Python.

 Specifically, we contribute literate execution as a paradigm, which we describe through motivating examples in \Secref{examples}. To support this work, we contribute new syntax and semantics for doc
comments in Fluid, and describe informally how literate execution views are derived from the dynamic dependence graph for a program (\Secref{implementation}). Finally, we reflect on the current state of play of our design and identify directions for future work in \Secref{evaluation}. Through this work, we aim to lower the barrier for authors to create transparent computational artefacts, making it
easier to communicate complex computational ideas through interactive documents.

\section{Motivating Examples}
\label{sec:examples}

Traditional literate programming interleaves static code with natural language commentary. Literate execution
inverts things, embedding the natural language commentary into the program's execution, so the program itself
participates in its own explanation. We illustrate this with two examples, first describing each from the reader's perspective (\S2.1), then the Fluid mechanisms that support these interactions (\S2.2).

\subsection{Two Reader Experiences}

\paragraph{Example 1: Matrix convolution.}

\figref{convolution-intermediates} shows a 5$\times$5 input image and a 3$\times$3 filter (A) producing a
5$\times$5 output via convolution. The reader starts by seeing these three matrices. Clicking on an output
cell establishes a \emph{persistent selection} (B), highlighting in green both the cell and the input cells in
the filter and image that contributed to it. The central column then reveals the 3$\times$3 intermediate
matrix (F) of neighbouring values that was averaged to produce the selected cell, together with a short
paragraph explaining the intermediate's role. Hovering over a cell instead produces a \emph{transient
selection} (C), highlighting related cells in blue (D) via the same dependency trace. Clicking additional
output cells surfaces additional intermediates side by side; hovering over an intermediate highlights the
input cells that fed into it and the output cell it contributes to. In this way a single executable definition
of convolution becomes an interactive teaching tool, inviting readers to form and test hypotheses about the
algorithm's behaviour.

\paragraph{Example 2: Transparent climate reporting.}

Figure~\ref{fig:ccra-nap3-intermediate} is adapted from Figure 1 of the UK Climate Change Committee's (CCC)
Independent Assessment of the Third National Adaptation Programme (NAP3). NAP3 outlines the UK Government's
adaptation actions for the period 2023--2028~\cite{climatechangecommittee2024}, developed in response to the
Third Climate Change Risk Assessment (CCRA3), which identified 61 risks and opportunities from climate change
in the UK; the independent assessment evaluates the extent to which NAP3 commitments address the gaps
highlighted in CCRA3. The reader sees a stacked bar chart summarising NAP3 actions assigned to Defra by
chapter, grouped by evaluation score, alongside a partial view of the underlying action records (showing
fields such as Action Owner and Evaluation score). Hovering over a bar segment updates the view \emph{in
situ}: the corresponding rows in the table are highlighted, and an explanatory paragraph appears in the
bottom-left panel describing the CCRA3 risk descriptors represented by those actions. The table thus serves as
a window onto the full dataset, letting the reader locate the relevant actions and risk descriptors for a
segment without manually browsing; the doc comment explains how the particular segment was constructed.

\subsection{Language Support for Literate Execution}

Both examples are built using two new Fluid features introduced in this paper: \emph{first-class doc comments}
and \emph{intermediate-value exploration}.

\paragraph{Doc comments.}

\begin{figure}
  \begin{subfigure}[t]{\textwidth}
    \begin{minipage}[t]{1.0\textwidth}
      \vspace{0pt}
      \includegraphics[width=\textwidth]{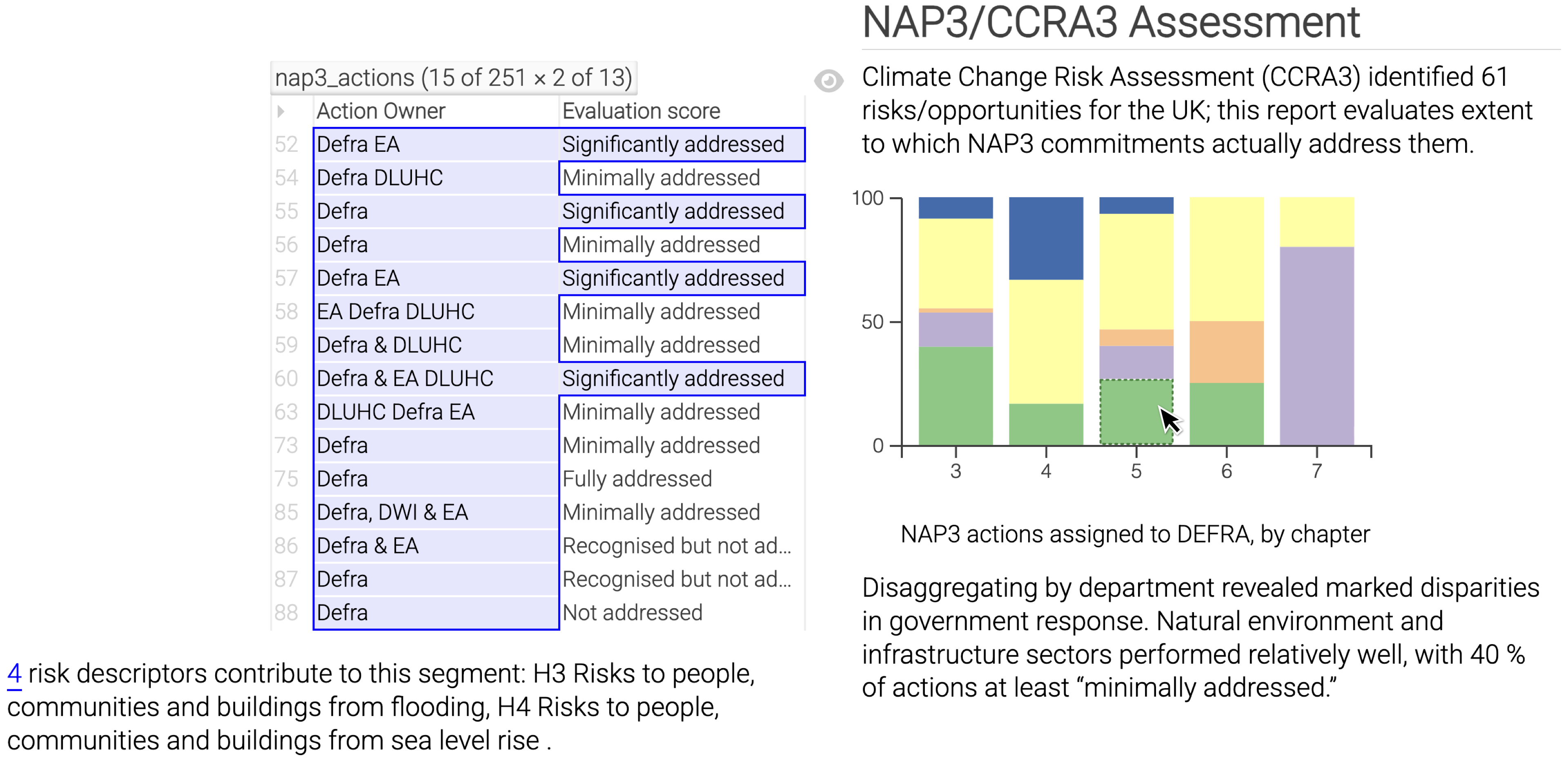}
    \end{minipage}%
  \end{subfigure}
  \vspace{1mm}
  \hrule
  \vspace{1mm}
  \begin{subfigure}[t]{\textwidth}
    \hspace{0mm}\includegraphics[width=1.0\textwidth]{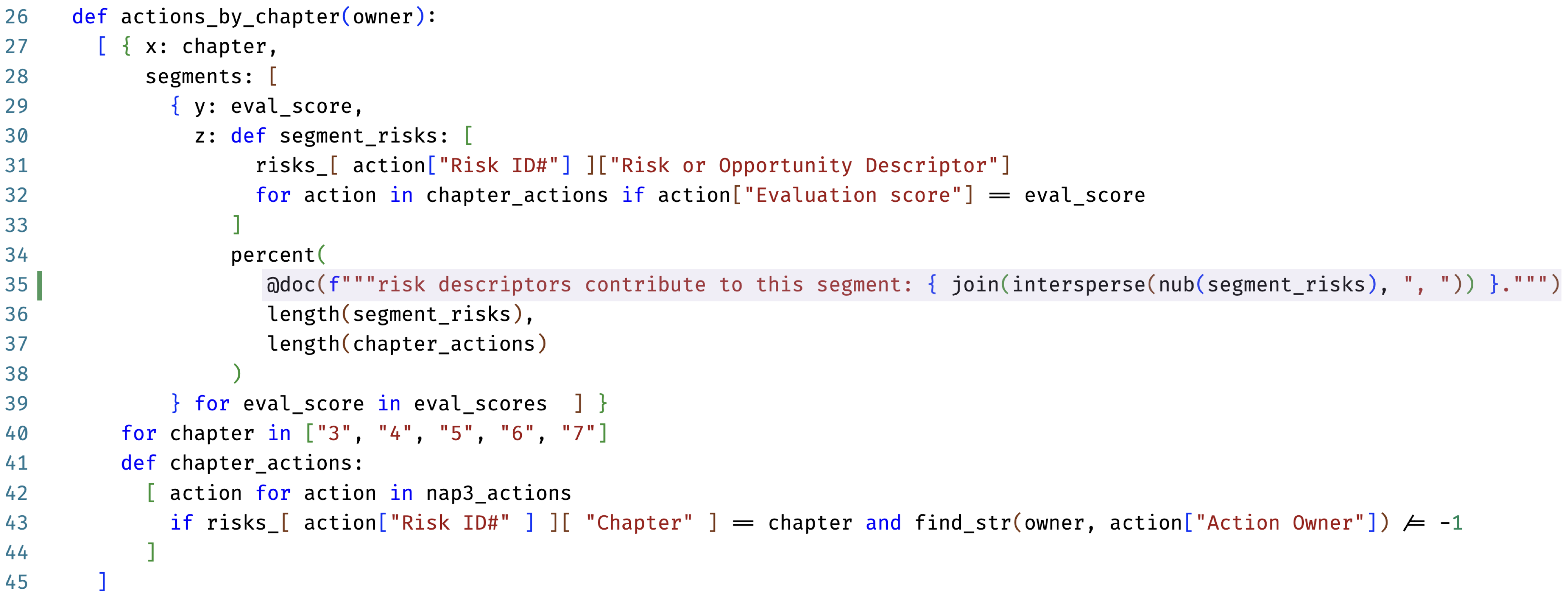}
  \end{subfigure}
  \vspace{-1em}
  \caption{Transparent climate reporting example}
  \label{fig:ccra-nap3-intermediate}
    \vspace{-1em}
\end{figure}

A doc comment is a dynamically computed expression of the form \kw{@doc(e)} that the user attaches to a
\emph{target expression}, syntactically by placing the \kw{@doc(e)} expression before the target expression,
with parentheses if disambiguation is needed. The doc expression $e$ must be of type \kw{Paragraph}. When the
target expression is evaluated, $e$ is evaluated at the same time, and the resulting paragraph is attached to
the value of the target expression as \emph{runtime documentation}. These paragraphs, which may contain
visualisations as well as text, can then be integrated into a visualisation of the running program. Example 1
attaches a doc comment to the matrix comprehension that produces each intermediate neighbourhood (E in
\figref{convolution-intermediates}). Example 2 attaches one to the intermediate computation that constructs
each bar segment (line 35 of \figref{ccra-nap3-intermediate}, bottom); its content is computed in the same
execution context as the segment itself, collecting the risk descriptors associated with the contributing
actions, removing duplicates with \kw{nub}, formatting the list with \kw{intersperse}, and concatenating with
\kw{join} to form the explanatory paragraph.

\paragraph{Intermediate-value exploration.}

A \kw{@doc} decorator marks the value its target expression computes as an observable \emph{intermediate},
surfaced to the reader during the interactions sketched above. When the reader selects an element of the
visualisation --- an output cell in Example 1, a bar segment in Example 2 --- the system traces through the
dynamic dependence graph to find the values that contributed to that element; any \kw{@doc}-tagged expressions
along the way yield intermediates, which are displayed alongside the selection together with their
documentation (F in \figref{convolution-intermediates} for Example 1; the bottom-left panel in
\figref{ccra-nap3-intermediate} for Example 2). The key invariant is that the same dependence graph that
drives the visualisation also drives the accompanying explanation: the selection on the aggregate, the
contributing records, and the computed text are kept in step automatically.

\section{Implementation}
\label{sec:implementation}

A key design choice for our system is to build on a new programming language, \OurLang, which provides
built-in support for dependency tracking and literate execution. This language-based approach is reflected in
some other literate tools; Idyll~\cite{conlen2018} develops a small DSL for authoring interactive documents,
while Literate Visualization~\cite{wood2019} builds on top of markdown with a sprinkling of Elm-based code
snippets for constructing Vega visualisations. Whereas with mainstream languages, the design is essentially
fixed, niche languages and DSLs give more flexibility, but also potentially present barriers to adoption. We
chose the latter approach to explore the design space more freely; future work could explore how existing languages and ecosystems might support literate execution.

We also made a surface-level choice for this particular work: to shift Fluid's syntax from its original
ML-inspired style~\cite{milner1997} to a more Pythonic form, potentially lowering the barriers to entry for
our target audience of data scientists and other domain experts unfamiliar with ML-like languages. This trades
some of the concision and idiomatic support for higher-order functional programming in favour of familiarity.
The change appears to have substantially improved the accessibility of the language for such programmers, though a rigorous evaluation is left to future work. We now describe the syntax
(\Secref{implementation:syntax}), semantics (\Secref{implementation:semantics}), and UI/interaction
implementation (\Secref{implementation:execution}) in more detail.

\subsection{Pythonic Syntax}
\label{sec:implementation:syntax}

The core of the new syntax is given in \figref{syntax}. The abstract syntax and core semantics of Fluid is
unchanged (from \citet{bond2025}), apart from the addition of \kw{@doc} comments; we will only provide
semantics for \kw{@doc}. A deeper treatment of the semantics is left to Appendix~\ref{appendix:semantics}.

The syntax is broadly similar to Python, with a few differences. Variables are not defined by
simply creating a name and assigning it a value, instead programmers must use the \kw{def} construct. Indeed,
owing to the similarity in the core language to ML, \OurLang is entirely expression-based; there are no
statements of any kind. This means that there is no direct assignment -- redefining a variable shadows it --
and we do not provide constructs such as return, break or continue. Finally, there are no classes: \OurLang
does not support an object system, nor does it support global state. Instead these restrictions guide programs
to be written as purely functional transformations of input data, and we exploit the graphical structure of
these transformations to produce our visualisations.

One of the points in which our syntax diverges somewhat from Python is in paragraph literals
$\exParagraph{\blockSeq{w}}$. These are composed of sequences of paragraph elements $w$, and are delimited by
$\exParagraph{...}$, similar to Python's mechanism for a triple quoted string. A paragraph element is either a
token $\paraToken{t}$, which corresponds to a string literal, or an unquoted term $\paraUnquote{s}$, which
corresponds to a term (expression) that gets evaluated and spliced back in to the paragraph so that the whole
item can be represented as a string once evaluated.

\begin{figure}
   {\small
   \begin{minipage}[t]{0.48\textwidth}
   \hspace{-3mm}%
   \begin{tabularx}{\textwidth}{rL{3cm}L{3cm}}
      &\textbf{Module or prog}&
      \\
      $m ::=$
      &
      $\blockSeq{d}$
      &
      module
      \\
      &
      $\blockSeq{d}\;s$
      &
      program
      \\[2mm]
      &\textbf{Definition}&
      \\
      $d ::=$
      &
      $\exDef{p}{s}$
      &
      variable
      \\
      &
      $\clauseFun{x}{\seq{p}}{s}$
      &
      function clause
      \\[2mm]
      &\textbf{Surface term}&
      \\
      $s ::=$
      &
      $x$
      &
      variable
      \\
      &
      $n$
      &
      int or float literal
      \\
      &
      $\exParagraph{\blockSeq{w}}$
      &
      paragraph literal
      \\
      &
      $\exApp{s}{\seq{s}}$
      &
      function call
      \\
      &
      $\exConstr{c}{\seq{s}}$
      &
      constructor
      \\
      &
      $\exDict{\seq{\bind{x}{s}}}$
      &
      dictionary
      \\
      &
      $\exProject{s}{x}$
      &
      lookup
      \\
      &
      $\exDProject{s}{s'}$
      &
      dynamic lookup
      \\
      &
      $\exFirstClassOp{\oplus}$
      &
      first-class operator
      \\
      &
      $\exBinaryApp{s}{\oplus}{s'}$
      &
      infix operator
      \\
      &
      $\exBinaryApp{s}{\exInfixFun{x}}{s'}$
      &
      infix function
      \\
      &
      $\exIfThenElse{s}{s_1}{s_2}$
      &
      if
      \\
      &
      $\exMatch{s}{\nonEmpty{\blockSeq{\mu}}}$
      &
      match
      \\
      &
      $\exList{\seq{s}}$
      &
      list
      \\
      &
      $\exListComp{s}{\nonEmpty{\seq{q}}}$
      &
      list comprehension
      \\
      &
      $\exFun{\seq{p}}{s}$
      &
      lambda
      \\
      &
      $\exDoc{s}{s'}$
      &
      doc expression
      \\
      \end{tabularx}
   \end{minipage}%
   \hspace{5mm}
   \begin{minipage}[t]{0.47\textwidth}
      \begin{tabularx}{\textwidth}{rL{2.8cm}L{2.9cm}}
      &\textbf{Match clause}&
      \\
      $\mu ::=$
      &
      $\clauseMatch{p}{s}$
      &
      \\[2mm]
      &\textbf{Pattern}&
      \\
      $p ::=$
      &
      $\pattVar{x}$
      &
      variable
      \\
      &
      $\pattRecord{\seq{\bind{x}{p}}}$
      &
      dictionary
      \\
      &
      $\pattConstr{c}{\seq{p}}$
      &
      constructor
      \\
      &
      $\pattList{\seq{p}}$
      &
      list
      \\[2mm]
      &\textbf{Qualifier}&
      \\
      $q ::=$
      &
      $\qualGuard{s}$
      &
      guard
      \\
      &
      $\qualDeclaration{p}{s}$
      &
      variable def
      \\
      &
      $\qualGenerator{p}{s}$
      &
      generator
      \\[2mm]
      &\textbf{Paragraph element}&
      \\
      $w ::=$
      &
      $\paraToken{t}$
      &
      token
      \\
      &
      $\paraUnquote{s}$
      &
      unquote
      \\[2mm]
      &\textbf{Comma sequence}&
      \\
      $\seq{z} ::=$
      &
      $\seqEmpty$
      &
      empty
      \\
      &
      $z$
      &
      singleton
      \\
      &
      $z\comma\;\seq{z}$
      &
      cons
      \\[2mm]
      &\textbf{Sequence}&
      \\
      $\blockSeq{z} ::=$
      &
      $\seqEmpty$
      &
      empty
      \\
      &
      $z\;\blockSeq{z}$
      &
      cons
      \\[7mm]
      \end{tabularx}
   \end{minipage}
   }
   \caption{Fluid's new Pythonic surface syntax}
   \label{fig:syntax}
\end{figure}

\subsection{Semantics of \kw{@doc} Comments}
\label{sec:implementation:semantics}
\begin{figure}
     {\flushleft \shadebox{$s \desugar e$}\hfill}
     \vspace{-1.85em}
   \begin{smathpar}
      \inferrule*[lab={\ruleName{$\desugar$-doc}}]
                 {
                   s \desugar e
                   \\
                   s' \desugar e'
                 }
                 {
                   \exDoc{s}{s'} \desugar \exDocCore{e}{e'}
                 }
      \and
      \inferrule*[lab=\ruleName{$\desugar$-paragraph}]
      {
         \blockSeq{w} \desugar \blockSeq{e}
      }
      {
         \exParagraph{\blockSeq{w}} \desugar \exConstr{\cPara}{\blockSeq{e}}
      }

   \end{smathpar}
   \\
   {\flushleft \shadebox {$\blockSeq{w} \desugar e$}
   \hfill \phantom{blah}}
   \begin{smathpar}
      \inferrule*[lab={\ruleName{$\desugar$-unquote}}]
      {
         s \desugar e
         \\
         \blockSeq{w} \desugar e'
      }
      {
         \paraUnquote{s}\cons \blockSeq{w} \desugar \exConstr{\cCons}{e, e'}
      }
      \and
      \inferrule*[lab={\ruleName{$\desugar$-token}}]
      {
         \blockSeq{w} \desugar e'
      }
      {
         \paraToken{t} \cdot \blockSeq{w} \desugar \exConstr{\cCons}{\exConstr{\kw{Str}}{t}, e'}
      }
      \and
      \inferrule*[lab={\ruleName{$\desugar$-empty}}]
      {
         \strut
      }
      {
         \seqEmpty \desugar \cNil
      }
   \end{smathpar} \\[2mm]
      {\small \flushleft \shadebox{$\gamma, e, V, G \Rightarrow v, G'$}%
      \begin{smathpar}
         \inferrule
         {
            \gamma, e, V, G \evalR v, G \cup \inStar{V}{\alpha}
            \\
            \gamma, e', V, G \evalR v', G \cup \inStar{V}{\alpha'}
            \\
            \fresh{\alpha,\alpha'}{G}
         }
         {
            \gamma, \exDocCore{e}{e'}, V, G \evalR \exDocCore{v}{v'}, G \cup \inStar{V}{\alpha} \cup \inStar{V}{\alpha'}
         }
      \end{smathpar}
   }
   \\
   \caption{Relevant rules from the semantics for doc comments, including the auxiliary desugarings.}
   \label{fig:doc-semantics}
\end{figure}

We highlight just those parts of the operational semantics concerning doc comments in \figref{doc-semantics},
which is broken down into three parts.

Overall, the operational semantics of Fluid is defined for the core language, whose expressions are denoted
$e$ with canonical value forms (expressions that can no longer reduce) denoted $v$. A desugaring phase $s
\desugar e$ maps surface terms $s$ to core expressions $e$. In the core language, $\exDocCore{e}{e'}$ denotes
doc comments, and hence is the target of the desugaring of surface doc comments. A surface paragraph literal
is desugared into a core expression consisting of the $\kw{Paragraph}$ constructor applied to a list formed
from an intermediate judgment (second desugaring relation). Paragraph tokens $t$ desugar to strings in the
core language. Unquoted terms $\paraUnquote{s}$ desugar to the expression they contain, and are $\kw{Cons}$ed
into the list.

In the Fluid operational model, core programs are evaluated into a \emph{dynamic dependence graph} (DDG),
whose vertices are abstract addresses $\alpha, \beta$ associated to values and edges $(\alpha, \beta)$ denote
that the value associated to $\beta$ depends on that associated to $\alpha$.

The bottom judgement form governs evaluating core expressions in a big-step operational semantics style where
$\gamma, e, V, G \evalR v, G'$ gives the semantics of evaluating expression $e$ under an environment $\gamma$,
with a graphical context $V,G$ -- to be read as a set of active vertices, and the dynamic dependence graph
under construction. This evaluates to value $v$ and an updated graph $G'$.

We show just the case for documented expressions, which can be read as follows: an expression
$\exDocCore{e}{e'}$ is decomposed into two parts, the comment itself $e$, and the term to which it is attached
$e'$. We evaluate these parts in turn, and combine the results into a documented value, composing both of
their updates to the graph in parallel (the two premises to the topmost rule). The notation $G \cup
\inStar{V}{\alpha}$ here refers to growing $G$ with the set of edges described by $V \times \set{\alpha}$.

In Appendix~\ref{appendix:semantics}, we show how other terms which involve recursively computing
sub-expressions -- such as function application -- thread the graph sequentially through the evaluation of
sub-terms.

\subsection{Deriving the Literate Execution}
\label{sec:implementation:execution}

We now provide a brief account of how the system constructs literate executions. Certain datatypes
in \OurLang are equipped with built-in visualisations: matrices, tables (represented as lists of
dictionaries), primitive values, and the various chart types. When embedded into a web page, a Fluid program
is evaluated and shows its output value as a visualisation; in the NAP3 example, this is a \kw{MultiView}
containing a stacked bar chart and some paragraphs. Clicking an interactive element of this visualisation
reveals the relevant parts of the inputs used to compute the output. The specific parts of these views that
get picked out are computed via the DDG.

Conceptually, the DDG records the transitive dependencies between values in the program. Each interaction
induces a slice of the dynamic dependence graph, traversing edges reachable from the vertices associated with
the values in the user's selection. When traversing the DDG, if a vertex is encountered that corresponds to a
value with a \kw{@doc} attached, that value is considered to be an intermediate value of interest. When the
user performs one of the usual interactions with the output (e.g.~click or hover), we reveal any views of
intermediates reachable from the selected vertices. If those intermediates are revealed persistently (by
clicking on an output element), the user can then further transiently interact with those intermediates, by
hovering the mouse. Doing so reveals both the upstream and downstream dependencies of that intermediate value.
In this way, narrative exposition of the value can be combined with a view of its data dependencies, revealing
more information about what the intermediate value represents and how it has been used.

\begin{figure}[t]
  \centering
  \includegraphics[width=0.6\linewidth]{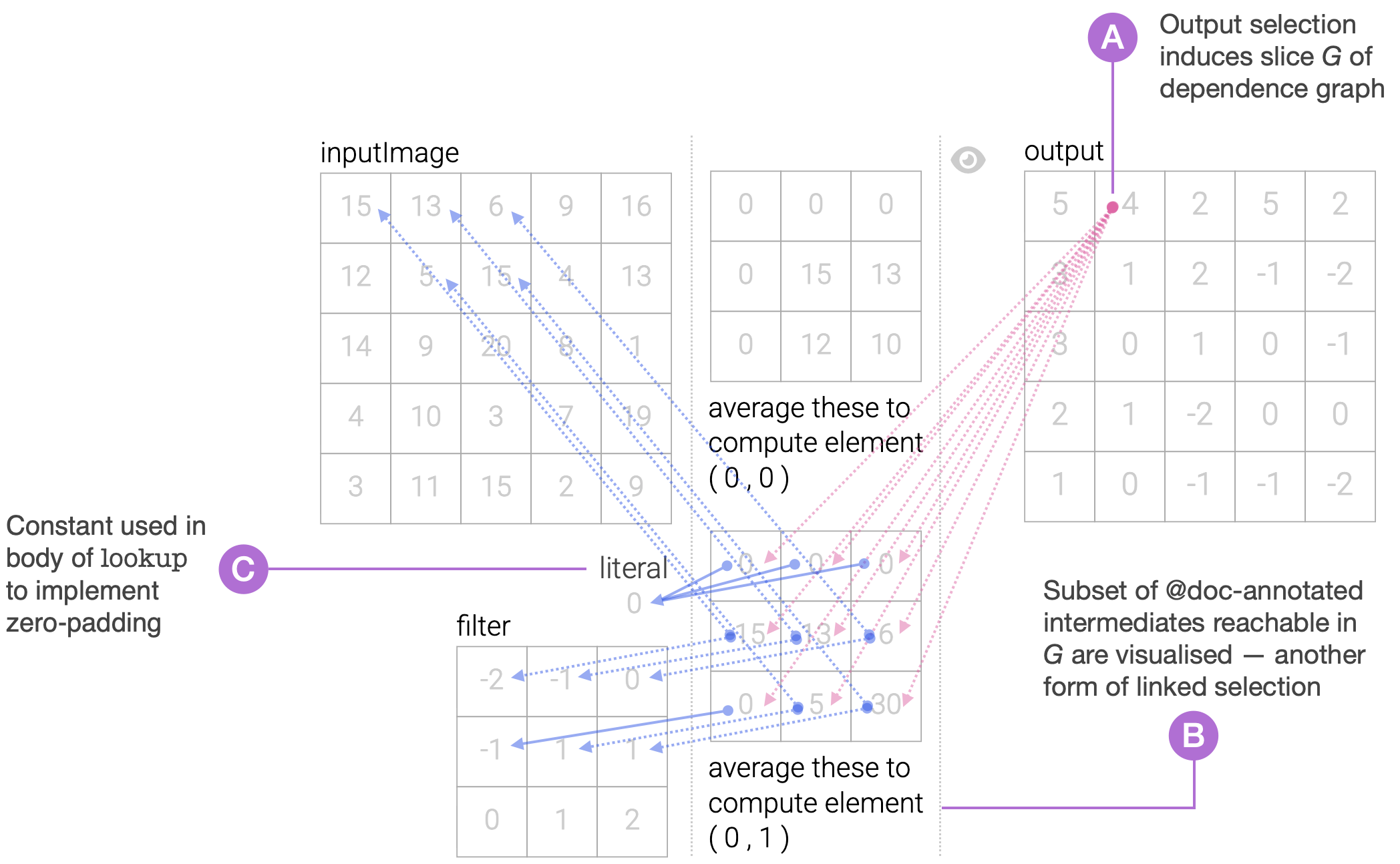}
  \caption{DDG slice induced by selecting an output cell in Example 1 (convolution). Pink dependency edges
  from the selected output (A) to the \kw{@doc}-tagged intermediate (B) arise via averaging; blue dependency
  edges from the intermediate to the inputs arise via multiplication.}
  \label{fig:ddg}
\end{figure}

\figref{ddg} shows this process concretely for Example 1. The reader has selected an output cell (A). Tracing
backwards through the DDG, the system encounters the \kw{@doc}-tagged matrix comprehension that produced the
3$\times$3 neighbourhood averaged to compute that output, and surfaces it as an intermediate (B) together with
its attached documentation. No further \kw{@doc} decorators are encountered along the backward traversal,
which terminates at the contributing input cells. Two details of the dependency analysis are worth noting.
First, (C) identifies a literal \kw{0} arising in the definition of \kw{lookup} that contributes to the output
because of \kw{lookup}'s role in implementing zero-padding at the image boundary. Although this constant is
(currently) not visualised in the UI, it is shown here to highlight the role it plays in the computation.
Second, each zero in the intermediate neighbourhood depends on only one input value (itself zero); these edges
are shown here using solid lines. The analysis recognises zero as the annihilator of multiplication, so there
is no dependency on the other argument. A zero produced by addition would necessarily depend on both summands.

\section{Reflection}
\label{sec:evaluation}

We now reflect on the current state of play of our design, focusing on two aspects: how documentation is attached to parts of the execution, and how the resulting literate execution is laid out for the reader.

\subsection{Inline vs. Out-of-Line Annotations}\label{sec:evaluation:inline}

Literate execution is an \emph{inline} approach to documentation, in contrast to approaches such as
\emph{literate tracing} \citep{sotoudeh2025}, where the documentation is maintained as an independent artefact; the split is somewhat analogous to the CSS/HTML separation of style and content. These two approaches offer
complementary advantages and disadvantages. With inline approaches, the documentation is situated in the code
context to which it applies, and as the code changes it is relatively straightforward to evolve the
documentation along with it. Out-of-line approaches like literate tracing decouple the secondary artifact from
the code, but suffer from the \emph{anchoring} problem: the difficulty of keeping references in the external
artifact in sync as the code changes. One mitigation is for the references to be query-driven rather than
bound to individual elements, providing more robustness to change; CSS selectors are the canonical example of
this approach. Another, suggested by recent work, is LLM-assisted maintenance of references across code
changes \cite{misback2025}; we plan to investigate these techniques as a way of mitigating the anchoring problem while retaining the flexibility of the out-of-line approach. The chief disadvantage of inline approaches is that annotations are tied to a single document perspective: producing a different literate document from the same codebase requires a fresh, unannotated copy of the code.

The same trade-off will apply to any layout metadata we might want to attach to \kw{@doc} elements in future,
such as layout hints or preferences, grouping, anchors tying paragraphs to specific interaction states, and so
on (discussed in the next subsection). Attaching such metadata to \kw{@doc} itself is the inline route:
presentation settings travel with content. A separate annotation layer would decouple the two, inheriting the
same trade-offs between direct references and query-driven alternatives discussed above.

\subsection{Layout of the Literate Document}\label{sec:evaluation:layout}

\paragraph{Current design.}

Our goal was to expose the intermediate values arising during a computation, allowing authors to attach
explanatory content to them in the design phase, and readers to interactively explore them in the deployed
document, giving end-users access to the internals of the computation, and allowing them to trace fine-grained
dependencies between inputs, intermediates, and outputs. As currently implemented, the author must use a
single mechanism for both exposing and documenting intermediates: the \kw{@doc} decorator, which
simultaneously selects an intermediate value for display and attaches a paragraph of documentation
(potentially with embedded visualisations and computations, via the paragraph-element splicing of
\Secref{implementation:syntax}). Which dependent values are to be considered \emph{inputs} is specified
slightly differently, via a separate configuration file; however, this separation is a contingent implementation detail rather than a deliberate division of concerns. Once specified, the set of candidate intermediates available to
a reader is fixed.

Moreover, beyond these basic hooks, most of the layout is currently outside the control of authors and
readers. The visualisation layer chooses how to render matrices, tabular data, and primitive values; there is
no mechanism to customise the rendering (e.g. grouping or sorting tables). Intermediates are arranged
vertically in a central column, with output to the right and inputs revealed to the left on selection. The
selection paradigm is also fixed: persistent by clicking, transient by hovering, with selection information
propagating downstream and upstream through the dependence graph automatically, highlighting any selected
parts of intermediates.

\paragraph{Flexible document layout.}

One richer layout paradigm we envisage is a \emph{layered intermediates} scheme: using the dependence graph to
select a topological ordering of intermediate values, presented as a kind of ``notebook-on-demand'', starting
with just a single cell (the output) and growing or shrinking progressively as the reader explores. This is
just one point in a larger space. Layout primitives that authors or readers might reasonably want include:
pinning an intermediate, hiding it, or choosing between different (but idiomatic) topological orderings. This
last feature might support different ``genres'' of interactive document: linear narratives, side-by-side
comparisons of two executions, step-by-step progressive reveals for teaching, dashboards, and no doubt others.

Most of these choices are candidates for either author or reader control: the reader often knows better than the author which arrangement suits their own reading, and for an artefact intended to be explored
interactively, that knowledge is often only available at read-time. Brushing and linking --- coordinated highlighting across views --- is one case in point.
In prior work on Fluid~\citep{perera2022}, a reader's selection on an input would additionally highlight any
\emph{related inputs}, and dually, a selection on an output would highlight any \emph{related outputs}.
Without intermediate values in the picture, the system could decide between these two modes automatically,
based on whether the reader clicked an input or an output. With \kw{@doc}-marked intermediates, however, each intermediate is both input and output, so the system can no longer decide automatically. For the present work we simply
disabled the feature; to re-enable it would require letting the reader toggle between those modes
interactively. In the notebook-on-demand setting, given an intermediate value $v$, a reader might want to see
the parts of $v$ which are related to a selection via some \emph{downstream} dependency one moment, and then
query for the parts of $v$ which are related to a selection via some \emph{upstream} dependency at a later
time. A coherent design would have to specify how author-supplied defaults compose with reader overrides, and
whether certain layout choices should remain fixed to ensure a consistent experience across readers.

The author/reader split itself may be less clean than it sounds. An author iterating on \kw{@doc} or layout
choices is already inhabiting the reader role, testing whether their choices serve the author's intent.
Going in the other direction, a natural extension adjacent to the visualisation-by-demonstration work cited in
\Secref{related} would let a reader interaction (revealing a particular group of intermediates and setting a
selection direction, say) be \emph{promoted} to an authored default. \emph{Debugging} moves fluidly between
authoring and reading: a developer asking \emph{why this output looked as it did} is both author and reader simultaneously, using the same \kw{@doc} and intermediate-value machinery to inspect their own code.
Fluid's runtime makes this connection tractable in a way that would be harder in, say, Python, where dependency tracking is not built into the runtime.

\section{Related Work}
\label{sec:related}

This work draws on several strands of prior work, including literate programming, tools for authoring
explanations, structured comments, and our own work on transparent programming languages. We discuss each in
turn.

\paragraph{Literate programming and its descendants}

Introduced by \citet{knuth1984}, \emph{literate programming} interleaves code and prose so that a program can
be read as both executable and explanatory text. Both literate programming and literate execution share the
structural principle of interleaving explanation with computational content, but they apply it to different
objects: literate programming explains the static structure of a program, while literate execution attempts to
explain its dynamic structure. \emph{Literate debugging} \cite{sugiyama2022} is a potentially complementary
idea, allowing specific execution points to be shared and reproduced via scripting and augmented with
user-provided contextualising documentation. \emph{Literate tracing} \cite{sotoudeh2025} also derives a
``partial'' or restricted view of an execution along with documentation that can mention values, albeit
without integrated provenance queries. Such work uses an \emph{out-of-band} approach where the metadata used
to derive the views is maintained as a separate document rather than embedded into the program, an approach
with a distinctive set of trade-offs that we would like to explore in future work. \emph{Literate
visualisation}~\cite{wood2019} applies similar ideas to the visualisation design process, guiding users to
document and explain each of their design decisions as they create visualisations.

\paragraph{Computational notebooks}

Computational notebooks such as Jupyter Notebooks~\cite{perez2007,kluyver2016} are perhaps the most widely
adopted descendant of literate programming, although the utilisation of literate features is
uneven~\cite{rule2018, wood2019}. While these environments are now the de facto standard for data science
work, they have well-documented limitations, such as those arising from their linear
structure~\cite{rule2018}, like disorganisation, lack of modularity, and challenges in unintentional
out-of-order execution~\cite{pimentel2019}. Notebooks provide a form of sequential decomposition, in that
cells expose intermediate steps, but they lack parallel decomposition: there is no provenance tracking to
connect individual parts of a structured output back to the specific inputs that produced them. Literate
execution aims to address both axes, attaching documentation and visualisations to arbitrary points in the
execution and using the dependency structure to allow readers to trace how individual parts of the output
relate to the input.

 \paragraph{Authoring tools for (explorable) explanations}

There is a substantial literature on explorable explanations, originating with \citet{victor2011a}. Much of
this work has focused on counterfactual transparency: allowing readers to adjust parameters, explore
alternative scenarios, and build intuition through interaction. Living Papers~\cite{heer2023} demonstrates how
computational papers can be automatically parsed, augmented, and redistributed as interactive, data-driven
documents, while Idyll~\cite{conlen2018} introduces a markup language (and subsequent structured editor-based
editing environment~\cite{hohman2020}) for authoring bespoke interactive essays. \citet{crichton2024} situate
this form of document-based language among a broader family of document languages, and \citet{hayatpur2025}
develop a DSL for authoring diagrams that illustrate algorithms. Our work is complementary: we propose that
provenance tracking provides a useful additional layer of infrastructure for \emph{decompositional}
transparency, enabling artefacts to become more self-explanatory by deriving interactions directly from the
semantics of the computation itself. Future work might explore how these two forms of transparency could be
integrated within document languages more generally.

Many tools explore augmenting text with visual explanations. \citet{masson2023a} develop a system that enables
users to add visual explanations to paragraphs of text, and \citet{zou2025} automatically generate word-scale
visualisations (\ala{} sparklines) to accompany text. A complementary line of work synthesises visualisation
specifications from user input: \citet{wang2019a,wang2021} develop `visualisation by demonstration', wherein
the user demonstrates a fragment of the intended visualisation and a complete specification is synthesised ---
drawing on prior works~\cite{baudel2006,kondo2014} based on direct manipulation. Literate execution shares the
goal of making text more informative through visual augmentation, but derives the augmentation from the
program's own structure rather than from hand-added visuals or user demonstrations.

\paragraph{Structured comments and embedded metadata}

Inline documentation systems like Javadoc~\cite{javadoc2025} and Doxygen~\cite{vanheesch2008} resemble
literate programming in that structured comments containing static references to program elements can be
rendered into separate documentation artefacts (e.g.~HTML pages).  These artefacts are statically generated,
separate from any program execution, and thus cannot report on intermediate computations.

Our approach shares some similarity also with Python's
docstrings~\cite{pep257} which make (certain forms of) documentation
available at runtime. A Python docstring is provided by a string
literal immediately following a function definition's head (see Figure~\ref{fig:python-doc}, left).
%
%
The function \pyinline{square} now has an additional attribute \pyinline{square.__doc__} (functions are
objects) which returns the docstring provided after the function head. Other built-in functions then leverage
this, e.g.~\pyinline{help(square)} uses \pyinline{square.__doc__} to produce a help message for this function.
Python docstrings however must be string literals, and cannot be `f-strings': formatted strings containing variables. For instance, \pyinline{f"""Square the input \{x\}"""} is not treated as a doc comment.

\begin{figure}[t]
	\vspace{-1em}
	\begin{subfigure}[t]{0.5\textwidth}
		\begin{python}
def square(x):
  """Squares the input parameter"""
  return (x**2)\end{python}
	\end{subfigure}%
	\begin{subfigure}[t]{0.5\textwidth}
		\begin{python}
@docformat
def square(x):
  """Square the input {x}"""
  return (x**2)\end{python}
	\end{subfigure}
	\vspace{-1em}
	\caption{Python example of docstrings (left) and a literate-execution
		style approach via a custom macro (right).}
	\label{fig:python-doc}
\end{figure}

Python's metaprogramming features can partially work around this limitation: a simple decorator (detailed in
Appendix~\ref{app:literate-exe-python}) can intercept function calls and format the docstring with the bound
argument values (Figure~\ref{fig:python-doc}, right), so that \pyinline{square(3)} prints `\texttt{Square the
input 3}' alongside returning the result. However, this is a narrow approximation of literate execution: it
substitutes values into text but does not connect those values to the surrounding computation. There is no
provenance tracking linking the documented function's arguments to the broader program's inputs, no mechanism
to surface intermediate values as separately explorable artefacts, and no integration with a visual
representation of the program that a reader can interact with.

Another related concept is that of ``doctests'', which allow executable input-output examples to be embedded
into documentation, usually in comments associated to a function.  Doctests are then detected by automated
testing tools, parsed as code, and evaluated as part of a test suite, e.g. in Python's
pytest~\cite{pytest-devteam2025} and Rust~\cite{rustdoc2025}. In our case, embedded code fragments are
usually visualisations used to present intermediate values, but one could imagine an idiomatic use where the
role of the visualisation is to validate and render some property of the value for testing or debugging. Such
approaches blur the line between standard program syntax and secondary notation, such as comments, which
usually have no semantic meaning and are essentially stripped out by a parser.

\paragraph{Execution visualisation}

A range of systems supports execution visualisation, overlapping with literate execution in different ways.
\citet{guo2013} developed Python Tutor, a teaching tool that visualises the step-by-step execution of Python
programs, allowing users to see the state of variables and data structures at each step. Literate execution
instead aims to support reader-driven traversal of the dependency graph with narrative documentation attached
to intermediate values, at the cost of requiring programs written in a provenance-tracking language; Python
Tutor works on mainstream Python with no instrumentation required.

\citet{lerner2020} presents \emph{projection boxes} (echoing earlier work by Victor~\cite{victor2012} and
Omnicode~\cite{kang2017}): reconfigurable inline displays of runtime values shown alongside code during live
programming. Projection boxes focus on the developer's experience while coding, displaying values at each
program point but not linking them to user selections or attaching narrative explanation. Literate execution
instead targets the reader of a deployed artefact, with the dependency graph driving interactions and
\kw{@doc} comments attaching prose; projection boxes in turn offer on-the-fly customisability to the
programmer that literate execution does not.

\citet{chen2022} present PI$_2$, which automatically generates fully functional interactive visualisation
interfaces from a sequence of SQL queries, analysing syntactic differences between queries to derive widgets
(sliders, dropdowns) and candidate visualisations. PI$_2$ addresses a complementary axis of transparency to
literate execution: \emph{counterfactual} (tweak a parameter, see a different chart) rather than
\emph{decompositional} (trace a value to its inputs). PI$_2$ is also fully automated (the author supplies only a query log), whereas literate execution requires hand-authored \kw{@doc} annotations.

\paragraph{Data provenance and transparent documents}

Language-based provenance-tracking has been explored for web languages~\cite{fehrenbach2016} and functional
languages~\cite{perera2012,ricciotti2017}. \citet{psallidas2018} were the first to explore the relationship
between data provenance and visual selections, allowing users to inspect data dependencies by selecting visual
elements, for SQL-like workflows. \citet{perera2022} developed a similar approach for a general-purpose
language, and gave a formal account of brushing and linking in terms of conjugate dependency analyses.
Building on this work, \citet{bond2025} reimplemented this as a \emph{cognacy} query over a dynamic dependence
graph (DDG), with a significant performance gain.

More recently, \citet{piscitelli2025} use Fluid to implement \emph{transparent text}: interactive scholarly documents in which hovering over a text fragment reveals the underlying data via provenance queries, using the same interpolated paragraph notation described in \Secref{implementation:syntax}. Literate execution uses the same notation but allows computed text to be attached as metadata to other values, via \kw{@doc}. Their work also contributes an LLM-based authoring workflow that we do not address; \kw{@doc} annotations are written by hand.








\section{Conclusions and Future Work}
\label{sec:conclusion}

 This work introduces \emph{literate execution}, an approach to authoring transparent data-driven documents in
 which documentation is attached to intermediate values within a computation. Provenance tracking reveals
 dependency relationships not only between the overall inputs and outputs, but also between the intermediate
 artefacts that arise along the way, allowing readers to explore both the sequential structure and the
 fine-grained data flow of a computation. Literate execution thus provides two important forms of
 decompositional transparency, complementing the counterfactual transparency offered by existing explorable
 explanation tools.

Several avenues for future work remain. \Secref{evaluation} sketched several design directions we would like
to explore for author and reader control of the literate-execution interface. A formal user study is needed
to explore these various directions. Such a study would help evaluate designs and identify areas for
improvement, particularly as there are also non-trivial usability challenges, both for authors and end-users
of the final artifact, which remain unexplored. Two specific technical limitations of our dependency-tracking
infrastructure are also worth highlighting. First, a \kw{@doc} comment cannot currently refer to the value it
is attached to; introducing a \kw{self} keyword would allow this and enable, for example, custom
visualisations such as presenting an intermediate table as a histogram, but requires a small redesign of the
dependency-graph construction to permit two-stage value initialisation. Second, dependencies in \OurLang only
arise when (partial) values are \emph{constructed}: projection forms like field access, variable access and
matrix indexing do not, leading to ``graph slices'' that may omit values one would expect to see.
Pattern-matching workarounds exist but are not always available and require deeper Fluid experience than can
be expected of an author.

A more substantial change to the provenance infrastructure we are pursuing is to extend it with
``intensional'' metadata, so that a user can see the operations and data involved in computing an output
selection. Through these developments and further evaluation, we hope to refine and expand the concept of
literate execution, making it a practical tool for authors seeking to create transparent and interactive
computational documents.

\pagebreak
\bibliographystyle{PLATEAU-Reference-Format}
\bibliography{comp-sci.bib} 
\appendix
\clearpage
\section{Full Semantics}
\label{appendix:semantics}
Here we discuss the evaluation relations of the core language, which we achieve by desugaring elements of
the surface language presented in the main paper. This core language features a restricted collection of expressions.
Semantically, Fluid is a simple call-by-name functional language, but with Pythonic surface syntax, and a structure for pattern-matching,
which is similar to a triemap \cite{peytonjones2022}, which we call an eliminator. The language is further parameterised
by a map $\Phi$, which takes primitive operations to their underlying implementations. Crucially, the language constructs the graph
which we use to build our interactive interface in evaluation piece by piece, by parsing, desugaring and evaluating a succession of modules.
One key point we would like to highlight here is that in practice, Fluid's surface syntax is indentation sensitive. For that reason,
we will not be giving an account of the parser that accepts the ``concrete surface syntax'' here, instead restricting our attention to
the ``abstract surface syntax'' -- the AST produced by the parser.


\subsection{Desugaring and the Core language}
\label{appendix:semantics:desugaring}
\begin{figure}[b]
   {\small
   \begin{minipage}[t]{0.48\textwidth}
      \hspace{-3mm}%
      \begin{tabularx}{\textwidth}{rL{3cm}L{3cm}}
         &\textbf{Expression}&
         \\
         $e ::=$
         &
         $x$
         &
         variable
         \\
         &
         $n$
         &
         int or float literal
         \\
         &
         $\exApp{e}{e'}$
         &
         application
         \\
         &
         $\exConstr{c}{\seq{e}}$
         &
         constructor
         \\
         &
         $\exDict{\seq{\bind{x}{e}}}$
         &
         dictionary
         \\
         &
         $\exProject{e}{x}$
         &
         lookup
         \\
         &
         $\exDProject{e}{e'}$
         &
         dynamic lookup
         \\
         &
         $\exFirstClassOp{\oplus}$
         &
         first-class operator
         \\
         &
         $\exLambda{\sigma}$
         &
         function
         \\
         &
         $\exLet{x}{e}{e'}$
         &
         let
         \\
         &
         $\exLetRec{\rho}{e}$
         &
         recursive let
         \\
         &
         $\exDocCore{e}{e'}$
         &
         doc expression
         \\[2mm]
         &\textbf{Continuation}&
         \\
         $\kappa ::=$
         &
         $e$
         &
         expression
         \\
         &
         $\sigma$
         &
         eliminator
      \end{tabularx}
   \end{minipage}%
   \hspace{3mm}
   \begin{minipage}[t]{0.47\textwidth}
      \begin{tabularx}{\textwidth}{rL{2.8cm}L{2.9cm}}
         &\textbf{Eliminator}&
         \\
         $\sigma ::=$
         &
         $\elimVar{x}{\kappa}$
         &
         variable
         \\
         &
         $\elimDict{\seq{x}}{\kappa}$
         &
         dictionary
         \\
         &
         $\elimConstr{\seq{\elimBind{c}{\kappa}}}$
         &
         constructor
         \\[2mm]
         &\textbf{Value}&
         \\
         $v ::=$
         &
         $\annot{v}{\alpha}$
         &
         annoted value
         \\
         &
         $n$
         &
         integer
         \\
         &
         $\exDict{\seq{\bind{x}{v}}}$
         &
         record
         \\
         &
         $\exConstr{c}{\seq{v}}$
         &
         constructor
         \\
         &
         $\exClosure{\gamma}{\rho}{\sigma}$
         &
         closure
         \\[2mm]
         &\textbf{Environment} &
         \\
         $\gamma ::=$
         & $\set{\seq{\bind{x}{v}}}$
         \\[2mm]
         &\textbf{Recursive definitions}
         \\
         $\rho ::=$
         & $\set{\seq{\bind{x}{\sigma}}} $
      \end{tabularx}
   \end{minipage}
   }
   \caption{Syntax of the core language, including values labeled with addresses}
   \label{fig:appendix:syntax-core}
\end{figure}

The relation $\desugar$ desugars the abstract surface language to the core language.

The core language is a restricted subset of the surface language, and is the language for which we will provide an account of
semantics. The first thing to note is that despite the Pythonic surface syntax, Fluid's core language has syntax in an ML-like style.
Therefore we write the desugared form of $\exDef{x}{s}$ terms as $\exLetTL{x}{e}$ for top level definitions or $\exLet{x}{e}{e'}$ for
definitions located within a larger definition. Another noteworthy point is the distinction between $\exFun{x}{s}$ and $\exLambda{\sigma}$.
The first is the definition of an anonymous function within the surface language, whilst the second is the representation of a lambda term
in the core language. The surface language has list comprehensions, which we desugar to a series of nested applications,
applied to lists created from generators, and then concatenate them using the \kw{concatMap}, which operates as it does in Haskell.
Doc comments have an explicit representation in the core syntax, written
$\exDocCore{e}{e'}$.

We show the core language in \figref{appendix:syntax-core}, and the rules for desugaring in \figref{appendix:desugar-term}.

\begin{figure}
   {\flushleft \shadebox{$s \desugar e$}\hfill}
   \begin{smathpar}
      \inferrule*[lab={\ruleName{$\desugar$-lambda}}]
      {
         s \desugar e
         \\
         \sigma \equal \elimBind{x}{e}
      }
      {
         \exFun{x}{s} \desugar \exLambda{\sigma}
      }
      \and
      \inferrule*[lab={\ruleName{$\desugar$-binary-apply}}]
      {
         s_1 \desugar e_1
         \\
         s_2 \desugar e_2
      }
      {
         \exBinaryApp{s_1}{\oplus}{s_2}
         \desugar
         \exApp{\exApp{\oplus}{e_1}}{e_2}
      }
      \and
      \inferrule*[lab={\ruleName{$\desugar$-fun-clauses}}]
      {
         \forall 1 \leq i \leq n \; (\clauseFun{x}{\seq{p}}{s} \desugar \sigma_i)_i
         \\
         \seq{\sigma} \desugar \sigma
      }
      {
         \seq{\clauseFun{x}{\seq{p}}{s}}
         \desugar
         \exLetRec{x}{\exLambda{\sigma}}
      }
      \and
      \inferrule*[lab={\ruleName{$\desugar$-match}}]
      {
         s \desugar e
         \\
         \mu \desugar\sigma
      }
      {
         \exMatch{s}{\mu}
         \desugar
         \exApp{\exLambda{\sigma}}{e}
      }
      \and
      \inferrule*[lab={\ruleName{$\desugar$-non-empty-list}}]
      {
         \seq{s} \equal s \cdot \seq{s}'
         \\
         s \desugar e
         \\
         \seq{s}' \desugar e'
      }
      {
         \exList{\seq{s}}
         \desugar
         \exConstr{\cCons}{e, e'}
      }
      \and
      \inferrule*[lab=\ruleName{$\desugar$-paragraph}]
      {
         \blockSeq{w} \desugar \blockSeq{e}
      }
      {
         \exParagraph{\blockSeq{w}} \desugar \exConstr{\cPara}{\blockSeq{e}}
      }
      \and
      \inferrule*[lab={\ruleName{$\desugar$-if}}]
      {
         s \desugar e
         \\
         s_1 \desugar e_1
         \\
         s_2 \desugar e_2
         \\
         \sigma = \elimConstr{\elimBind{\cTrue}{e_1}, \elimBind{\cFalse}{e_2}}
      }
      {
         \exIfThenElse{s}{s_1}{s_2}
         \desugar
         \exApp{\exLambda{\sigma}}{e}
      }
      \and
      \inferrule*[lab={\ruleName{$\desugar$-list-comp-done}}]
      {
         s \desugar e
      }
      {
         \exListComp{s}{\seqEmpty}
         \desugar
         \exConstr{\cCons}{e, \cNil}
      }
      \and
      \inferrule*[lab={\ruleName{$\desugar$-list-comp-guard}}]
      {
         s \desugar e
         \\
         \exListComp{s'}{\seq{q}} \desugar e'
         \\
         \sigma = \elimConstr{\elimBind{\cTrue}{e'}, \elimBind{\cFalse}{\cNil}}
      }
      {
         \exListComp{s'}{\qualGuard{s} \cons \seq{q}}
         \desugar
         \exApp{\exLambda{\sigma}}{e}
      }
      \and
      \inferrule*[lab={\ruleName{$\desugar$-list-comp-decl}}]
      {
         s \desugar e
         \\
         (\clause{p}{\exListComp{s'}{\seq{q}}}) \desugar \sigma
      }
      {
         \exListComp{s'}{\qualDeclaration{p}{s} \cons \seq{q}}
         \desugar
         \exApp{\exLambda{\sigma}}{e}
      }
      \and
      \inferrule*[lab={\ruleName{$\desugar$-list-comp-gen}}]
      {
         s \desugar e
         \\
         (\clause{p}{\exListComp{s'}{\seq{q}}}) \orElse \mu
         \\
         \mu \desugar \sigma
      }
      {
         \exListComp{s'}{\qualGenerator{p}{s} \cons \seq{q}}
         \desugar
         \exApp{\exApp{\varConcatMap}{\exLambda{\sigma}}}{e}
      }
      \and
      \inferrule*[lab={\ruleName{$\desugar$-doc}}]
                 {
                   s \desugar e
                   \\
                   s' \desugar e'
                 }
                 {
                   \exDoc{s}{s'} \desugar \exDocCore{e}{e'}
                 }
   \end{smathpar}
   \\[2mm]
   {\flushleft \shadebox {$\blockSeq{w} \desugar e$}
   \hfill \phantom{blah}}
   \begin{smathpar}
      \inferrule*[lab={\ruleName{$\desugar$-unquote}}]
      {
         s \desugar e
         \\
         \blockSeq{w} \desugar e'
      }
      {
         \paraUnquote{s}\cons \blockSeq{w} \desugar \exConstr{\cCons}{e, e'}
      }
      \and
      \inferrule*[lab={\ruleName{$\desugar$-token}}]
      {
         \blockSeq{w} \desugar e'
      }
      {
         \paraToken{t} \cdot \blockSeq{w} \desugar \exConstr{\cCons}{\exConstr{\kw{Str}}{t}, e'}
      }
      \and
      \inferrule*[lab={\ruleName{$\desugar$-empty}}]
      {
         \strut
      }
      {
         \seqEmpty \desugar \cNil
      }
   \end{smathpar}
   \\[2mm]
   {\flushleft \shadebox{$l \desugar e$}
   \hfill \phantom{blah}}
   \begin{smathpar}
      \inferrule*[lab={\ruleName{$\desugar$-list-rest-nil}}]
      {
         \strut
      }
      {
         \exListEnd
         \desugar
         \exListEnd,
         \cNil
      }
      \and
      \inferrule*[lab={\ruleName{$\desugar$-list-rest-cons}}]
      {
         s \desugar e
         \\
         l \desugar e'
      }
      {
         (\exListNext{s}{l})
         \desugar
         \exConstr{\cCons}{e, e'}
      }
   \end{smathpar}
   \\[2mm]
   \begin{minipage}[t]{0.48\textwidth}
      {\small \flushleft \shadebox{$\gamma \desugar \gamma'$}
      \begin{smathpar}
         \inferrule*[lab={\ruleName{$\desugar$-env}}]
         {
            \seq{v} \desugar \seq{v}'
         }
         {
            \set{\seq{\bind{x}{v}}} \desugar \set{\seq{\bind{x}{v}}'}
         }
      \end{smathpar}
      }
   \end{minipage}%
   \hspace{2mm}%
   \begin{minipage}[t]{0.48\textwidth}
      {\small \flushleft \shadebox {$u \desugar v$}
         \begin{smathpar}
            \inferrule*[lab={\ruleName{$\desugar$-closure}}]
            {
                  \gamma \desugar \gamma'
                  \\
                  \seq{\mu} \desugar \seq{\sigma}
                  \\
                  \mu \desugar \sigma
            }
            {
                  \exClosure{\gamma}{\seq{\bind{x}{\mu}}}{\mu} \desugar \exClosure{\gamma'}{\seq{\bind{x}{\sigma}}}{\sigma}
            }
         \end{smathpar}
      }
   \end{minipage}
   \caption{Desugaring for terms, environments and closures}
   \label{fig:appendix:desugar-term}
\end{figure}

Paragraphs are interesting, since they are represented in the surface language as a special kind of literal, with unquoted expressions embedded
within, but in the core language we represent them as a datatype, \kw{Paragraph}, which contains a list of elements. These elements exploit
\OurLang's dynamic typing, they can be of heterogeneous type, provided they have built-in visualisations.

Function clauses are combined together into lambdas in the desugaring pass. The rules for how we do this are expressed in
\figref{appendix:desugar-clauses}. The key structure here, denoted with the variable $\sigma$ is that of the \textit{eliminator}, mentioned
above. Multiple function clauses with the same name are combined into a single eliminator, and this will be pattern matched against
in evaluation later. We let $k$ range over $(\seq{p}, p' = s)$ pairs. The first component is the currently active stack of
sub patterns, active during the processing of a top-level pattern - an argument to a function or Constructor, or as part of a
list comprehension. It starts with just $p$ and ends up empty. The second component
stores remaining top-level patterns and is empty except for in curried functions. We omit desugaring rules for the terms that are uninteresting
or obvious -- i.e. the terms which are structurally identical between the surface and core languages.

As convenient notation, we define the following, which comes in handy for piecewise function definitions:
\begin{definition}
   Suppose $k=(\seq{p}, p'\equal s)$.
   \begin{itemize}
      \item Define $p \clauseCons k \eqdef (p \cons \seq{p}, p' \equal s)$
      \item Define $\seq{p}^{\dagger} \clauseConcat k \eqdef (\seq{p}^{\dagger}\concat \seq{p}, p' \equal s)$
   \end{itemize}
   And the same notation can be extended to sequences.
\end{definition}

\begin{figure}
   {\flushleft \shadebox{$\mu \desugar \sigma$}
   \hfill}
   \begin{smathpar}
      \inferrule*[
         lab={\ruleName{$\desugar$-fun}}
      ]
      {
         \seq{(p, \clause{\seq{p}'}{s})}
         \desugar \sigma
      }
      {
         \seq{\clause{p \cons \seq{p}'}{s}}
         \desugar
         \sigma
      }
   \end{smathpar}
   \\[2mm]
   {\flushleft \shadebox{\strut $\seq{k} \desugar \kappa$}
   \hfill}
   \begin{smathpar}
      \inferrule*[
         lab={\ruleName{$\desugar$-clauses-done}}
      ]
      {
         s \desugar e
      }
      {
         (\seqEmpty, \clause{\seqEmpty}{s}) \desugar e
      }
      \and
      \inferrule*[
         lab={\ruleName{$\desugar$-clauses-next-arg}}
      ]
      {
         \seq{(p, \clause{\seq{p}}{s})}
         \desugar
         \sigma
      }
      {
         \seqRange{(\seqEmpty, \clause{p_1 \cons \seq{p}_1}{s_1})}
                  {(\seqEmpty, \clause{p_j \cons \seq{p}_j}{s_j})}
         \desugar
         \exLambda{\sigma}
      }
      \and
      \inferrule*[
         lab={\ruleName{$\desugar$-clauses-var}}
      ]
      {
         \seq{k} \desugar \kappa
      }
      {
         \seqRange{(\pattVar{x} \clauseCons k_1)}{(\pattVar{x} \clauseCons k_j)}
         \desugar
         \elimVar{x}{\kappa}
      }
      \and
      \inferrule*[
         lab={\ruleName{$\desugar$-clauses-dict}}
      ]
      {
         (\seqRange{\seq{p_1}}{\seq{p_j}}) \clauseConcat \seq{k} \desugar \kappa
      }
      {
         (\seqRange{\pattRecord{\seq{\bind{x}{p_1}}}}{\pattRecord{\seq{\bind{x}{p_j}}}})
         \clauseCons
         \seq{k}
         \desugar
         \elimDict{\seq{x}}{\kappa}
      }
      \and
      \inferrule*[
         lab={\ruleName{$\desugar$-clauses-constr}}
      ]
      {
         ((\seq{p}_i \clauseConcat k_i) \mid c_i = c) \desugar \kappa_c
         \\
         \datatype{c} = D
         \quad
         (\forall c \in \set{\seq{c}})
      }
      {
         \seq{\pattConstr{c}{\seq{p}} \clauseCons k}
         \desugar
         \elimConstr{\elimBind{c}{\kappa_c} \mid c \in \set{\seq{c}}}
      }
      \and
      \inferrule*[
         lab={\ruleName{$\desugar$-clauses-non-empty-list}}
      ]
      {
         \seq{p} \equal p \cons \seq{p}'
         \\
         (\pattConstr{\cCons}{p \cons \seq{p}'} \clauseCons k) \cons \seq{k}
         \desugar
         \kappa
      }
      {
         (\pattList{\seq{p}} \clauseCons k)
         \cons
         \seq{k}
         \desugar
         \kappa
      }
      \and
      \inferrule*[
         lab={\ruleName{$\desugar$-list-end}}
      ]
      {
         (\cNil \clauseCons k) \cons \seq{k}
         \desugar
         \kappa
      }
      {
         (\pattListEnd \clauseCons k)
         \cons
         \seq{k}
         \desugar
         \kappa
      }
      \and
      \inferrule*[
         lab={\ruleName{$\desugar$-list-cons}}
      ]
      {
         (\pattConstr{\cCons}{p \cons o} \clauseCons k) \cons \seq{k}
         \desugar
         \kappa
      }
      {
         (\pattListNext{p}{o} \clauseCons k)
         \cons
         \seq{k}
         \desugar
         \kappa
      }
   \end{smathpar}
   \caption{Desugaring for function definitions}
   \label{fig:appendix:desugar-clauses}
\end{figure}

\begin{figure}[h!]
   {\small \flushleft \shadebox{$k, \alpha \orElse \mu$}
   \begin{smathpar}
      \inferrule*[
         lab={\ruleName{$\orElse$-done}}
      ]
      {
         \strut
      }
      {
         (\seqEmpty, \clause{\seqEmpty}{s}), \alpha
         \orElse
         \clause{\seqEmpty}{s}
      }
      \and
      \inferrule*[
         lab={\ruleName{$\orElse$-var}}
      ]
      {
         k, \alpha \orElse \seqRange{k_1}{k_j}
      }
      {
         \pattVar{x} \clauseCons k, \alpha
         \orElse
         \seqRange{\pattVar{x} \clauseCons k_1}
                  {\pattVar{x} \clauseCons k_j}
      }
      \and
      \inferrule*[
         lab={\ruleName{$\orElse$-record}}
      ]
      {
         \seq{p} \clauseConcat k, \alpha
         \orElse
         \seqRange{\seq{p_1} \clauseConcat k_1}
                  {\seq{p_j} \clauseConcat k_j}
      }
      {
         \pattRecord{\seq{\bind{x}{p}}} \clauseCons k, \alpha
         \orElse
         \seqRange{\pattRecord{\seq{\bind{x}{p_1}}} \clauseCons k_1}
                  {\pattRecord{\seq{\bind{x}{p_j}}} \clauseCons k_j}
      }
      \and
      \inferrule*[
         lab={\ruleName{$\orElse$-constr}},
         right={$\ctrsF(D) = c \cons \set{\seq{c}}$}
      ]
      {
         \seq{p} \clauseConcat k, \alpha
         \orElse
         \seq{\seq{p} \clauseConcat k}
         \\
         \seq{k}' =
         ((\pattVar{\varAnon} \mid n \numleq \length{\seq{p}'}, \clause{\seqEmpty}{\annot{\cNil}{\alpha}})
         \mid
         i \numleq \length{\seq{c}})
         \\
         k = (\seq{p}', \clause{\seqEmpty}{s})
      }
      {
         \pattConstr{c}{\seq{p}} \clauseCons k, \alpha
         \orElse
         (\seqRange{\pattConstr{c}{\seq{p}_1} \clauseCons k_1}
                   {\pattConstr{c}{\seq{p}_j} \clauseCons k_j})
         \concat
         (\pattConstr{c_i}{\pattVar{\varAnon} \mid n \numleq \arity{c_i}}
         \clauseCons k'_i
         \mid i \numleq \length{\seq{c}})
      }
      \and
      \inferrule*[
         lab={\ruleName{$\orElse$-list}}
      ]
      {
         p \clauseCons \seq{o} \clauseCons k, \alpha
         \orElse
         \seq{p \clauseCons \seq{o} \clauseCons k}
         \\
         k = (\seq{p}', \clause{\seqEmpty}{s})
      }
      {
         (\exList{p \cons \seq{o}})
         \clauseCons k, \alpha
         \orElse
         \seqRange{(\pattList{p_1\cons \seq{o_1}}) \clauseCons k_1}
                  {(\pattList{p_j \cons \seq{o_j}}) \clauseCons k_j},
         \pattListEnd \cons \seq{p}', \clause{\seqEmpty}{\annot{\cNil}{\alpha}}
      }
   \end{smathpar}}
   \caption{Completing an (uncurried) clause with a default value of $\cNil$}
   \label{fig:appendix:or-else}
\end{figure}

\subsection{Evaluating Expressions}
\begin{figure}
   \vspace{-1em}
    {\small \flushleft \shadebox{$\gamma, e, V, G \evalR v, G'$}%
    \begin{smathpar}
       \inferrule*[
       lab={\ruleName{$\evalR$-var}}
       ]
       {
          \strut
       }
       {
          \gamma \cons (\bind{x}{v}), x, V, G
          \evalR
          v,
          G
       }
       \and
       \inferrule*[
          lab={\ruleName{$\evalR$-int}}
       ]
       {
          \fresh{\alpha}{G}
       }
       {
          \gamma, n, V, G
          \evalR
          \annInt{n}{\alpha},
          G \cup \inStar{V}{\alpha}
       }
       \and
       \inferrule*[lab={\ruleName{$\evalR$-function}}]
       {
          \fresh{\alpha}{G}
       }
       {
          \gamma, \exFun{\sigma}, V, G
          \evalR
          \annClosure{\gamma}{\varnothing}{\sigma}{\alpha},
          G \cup \inStar{V}{\alpha}
       }
       \and
       \inferrule*[lab={\ruleName{$\evalR$-dict}}]
       {
          \gamma, \seq{e}, V, G \evalR \seq{v}, G'
          \\
          \fresh{\alpha}{G'}
       }
       {
          \gamma, \exDict{\seq{\bind{x}{e}}}, V, G
          \evalR
          \annDict{\seq{\bind{x}{v}}}{\alpha},
          G' \cup \inStar{V}{\alpha}
       }
       \and
       \inferrule*[
          lab={\ruleName{$\evalR$-project}}
       ]
       {
          \gamma, e', V, G \evalR y, G'
          \\
          \gamma \vdash y : \kw{String}
          \\
          \gamma, e, V, G \evalR \annDict{\seq{\bind{x}{v}} \cons (\bind{y}{u})}{\alpha}, G''
       }
       {
          \gamma, \exDProject{e}{e'}, V, G
          \evalR
          u,
          G''
       }
       \\
       \inferrule*[lab={\ruleName{$\evalR$-constr}}
       , right={$\Sigma(c) = |\seq{e}|$}]
       {
          \gamma, \seq{e}, V, G \evalR \seq{v}, G'
          \\
          \fresh{\alpha}{G'}
       }
       {
          \gamma, \exConstr{c}{\seq{e}}, V, G
          \evalR
          \annConstr{c}{\seq{v}}{\alpha},
          G' \cup \inStar{V}{\alpha}
       }
       \and
       \inferrule*[ lab={\ruleName{$\evalR$-foreign-app}}
       , right={$\Phi(f) = |\seq{e}|$}]
       {
          \gamma, \seq{e}, V, G_1 \evalR \seq{v}, G_2
          \\
          \interpret{f}(\seq{v}, G_2) = (u, G_3)
       }
       {
          \gamma, \exApp{\exFirstClassOp{f}}{\seq{e}}, V, G_1
          \evalR
          u,
          G_3
       }
       \and
       \inferrule*[
          lab={\ruleName{$\evalR$-app}},
          width=5.5in,
       ]
       {
          \gamma, e, V, G_1 \evalR \annClosure{\gamma_1}{\rho}{\sigma}{\alpha}, G_2
          \\
          \gamma_1, \rho, \set{\alpha}, G_2 \closeDefs \gamma_2, G_3
          \\
          \gamma, e', V, G_3 \evalR v', G_4
          \\
          v', \sigma \match \gamma_3, e^\twoPrime, V'
          \\
          \gamma_1 \concat \gamma_2 \concat \gamma_3, e^\twoPrime,  V' \cup \set{\alpha}, G_4 \evalR u, G_5
       }
       {
          \gamma, \exApp{e}{e'}, V, G_1
          \evalR
          u,
          G_5
       }
       \and
       \inferrule*[
          lab={\ruleName{$\evalR$-doc}}]
         {
            \gamma, e, V, G \evalR v, G \cup \inStar{V}{\alpha}
            \\
            \gamma, e', V, G \evalR v', G \cup \inStar{V}{\alpha'}
            \\
            \fresh{\alpha,\alpha'}{G}
         }
         {
            \gamma, \exDocCore{e}{e'}, V, G \evalR \exDocCore{v}{v'}, G \cup \inStar{V}{\alpha} \cup \inStar{V}{\alpha'}
         }
    \end{smathpar}}
    \\[-1cm]
    {\small \flushleft \shadebox{$\gamma, \seq{e}, V, G \evalR \seq{v}, G'$}%
    \begin{smathpar}
       \inferrule*[
          right={$n = \length{\seq{e}}$}
       ]
       {
          \gamma, e_i, V, G_i \evalR v_i, G_{i + 1}
          \\
          (\forall i \numleq n)
       }
       {
          \gamma, \seq{e}, V, G_1
          \evalR
          \seq{v},
          G_{n + 1}
       }
       \and
    \end{smathpar}}
    {\small \flushleft \shadebox{$\gamma, \rho, V, G \closeDefs \gamma', G'$}
    \begin{smathpar}
       \inferrule*[
          right={$n = \length{\seq{x}}$}
       ]
       {
          \gamma'(x_i) = \annClosure{\gamma}{\rho}{\rho(x_i)}{\alpha_i}
          \\
          \alpha_i \notin \dom{G_i}
          \\
          G_{i+1} = G_i \cup \inStar{V}{\alpha_i}
          \quad
          (\forall i \numleq n)
       }
       {
          \gamma, \rho, V, G_1
          \closeDefs
          \gamma',
          G_{n+1}
       }
    \end{smathpar}
    }
    \caption{Operational semantics with dependence graph}
    \label{fig:appendix:eval}
 \end{figure}

As mentioned above, evaluation proceeds in a simple call-by-name style, we present a big-step evaluation relation in \figref{appendix:eval}.
We read the judgement $\gamma, e, V, G \evalR v, G'$ as stating that the term $e$m under an environment $\gamma$, vertex set V, and
dependence graph $G$, evaluates to the value $v$ and an extended dependence graph $G'$. In this judgement, $V$ refers to the values
which have been consumed by the current function call. It starts off empty, and is filled by the root vertices of values consumed by function calls.
These vertices are then attached to new values created in the function body $e$, growing the dependence graph in star-shaped pieces.

Rules which introduce values such as integers, functions, and constructors, all assign a fresh address for the (partial) value under construction.
For example, in the evaluation rule for integers, the result value has the form $\annInt{n}{\alpha}$, indicating a fresh vertex $\alpha$ being added
to the graph, which is the target vertex for the edges in $\inStar{V}{\alpha}$, and likewise for other introduction forms.
Anonymous functions $\exLambda{\sigma}$ construct closures $\exClosure{\gamma}{\envEmpty}{\sigma}_\alpha$ with a fresh address, and capture the
current environment $\gamma$. The $\envEmpty$ is the set of mutually recursive definitions associated with the function, and $\sigma$ is the
functions eliminator, as well as extending the graph with $\inStar{V}{\alpha}$. Each vertex that
is added to the graph in this way is annotated with a reference to the value with that vertex at its root. \OurLang uses these annotations during interaction
to map from elements of a graph slice back to values. Crucially, this is what allows \OurLang to find values annotated with \kw{@doc}, since
an expression annotated with \kw{@doc} is evaluated like normal, but the value referenced by the graph also contains the evaluated comment. The runtime
traverses the graph, and for every vertex it encounters, can check if the value it points to has a comment. If it does, that value is automatically
considered an \emph{intermediate}, and visualised in the user interface.

Rules which recursively evaluate subterms take the graph under construction, and sequentially thread it through the evaluation of each subterm,
the impact of which is to ensure that $\alpha$'s assigned in each subterm are fresh with respect to the global context. A simple example of this
is the rule for evaluating sequences of expressions, $\gamma, \seq{e}, V, G \evalR \seq{v}, G'$, which evaluates each element in turn,
passing the updated graph to the next element. A more complicated example is shown in the rule \ruleName{$\evalR$-app}, where the function is
evaluated to a closure, the set of mutually recursive definitions $\rho$ are closed over ($\closeDefs$) to update the environment, the argument is evaluated, and matched
against, and finally the appropriate function clause is evaluated with the argument included in the environment.

Closing definitions is accomplished through the close-defs relation, $\gamma, \rho, V, G \closeDefs \gamma', G'$. This relation takes a set $\rho$
of recursive definitions, a vertex set $V$, and a graph $G$, returning an environment $\gamma'$ containing a closure $\exClosure{\gamma}{\rho}{\rho(x_i)}_{\alpha_i}$
for each definition $\rho_i \in \rho$, and an extended dependence graph, $G'$.

Values are consumed by pattern matching in function calls defined in \figref{appendix:pattern-matching}. Pattern matching is defined for data-type
constructors -- including syntactic sugar for values like tuples and lists -- as well as variables and dictionaries. Notably, a dictionary match
can be performed against a subset of key value pairs, and does not need to be for the entire dictionary. The pattern matching judgement
matches a sequence of values $\seq{v}$ against a continuation $\kappa$, returning the selected branch, as well as the vertices
of each value that was successfully matched against.

\begin{figure}
   {\small \flushleft \shadebox{$\seq{v}, \kappa \match \gamma, e, V$}\hfill
   \begin{smathpar}
      \inferrule*[
         lab={\ruleName{$\match$-done}}
      ]
      {
         \strut
      }
      {
         \seqEmpty, e \match \envEmpty, e,  \emptyset
      }
      \and
      \inferrule*[lab={\ruleName{$\match$-constr}}
                , right={$\Sigma(c) = |\seq{v}|$}]
      {
         \seq{v} \concat \seq{v}', \kappa
         \match
         \gamma, e, V
      }
      {
         \annot{\exConstr{c}{\seq{v}}}{\alpha} \cons \seq{v}', (\elimBind{c}{\kappa}) \cons \elimConstr{\seq{\elimBind{c}{\kappa}}}
         \match
         \gamma, e, \alpha \cons V
      }
      \\
      \inferrule*[lab={\ruleName{$\match$-var}}]
      {
         \seq{v}, \kappa \match \gamma, e, V
      }
      {
         v \cons \seq{v}, \elimVar{x}{\kappa}
         \match
         \gamma \cons (\bind{x}{v}), e, V
      }
      \and
      \inferrule*[
         lab={\ruleName{$\match$-dictionary}}
      ]
      {
         \exDict{\seq{\bind{y}{u}}} \subseteq \exDict{\seq{\bind{x}{v}}}
         \\
         \seq{u} \concat \seq{v}', \kappa
         \match
         \gamma, e, V
      }
      {
         \annot{\exDict{\seq{\bind{x}{v}}}}{\alpha} \cons \seq{v}',
         \elimDict{\seq{y}}{\kappa}
         \match
         \gamma, e, \alpha \cons V
      }
   \end{smathpar}}
\vspace{-0.5cm}
\caption{Pattern matching}
\label{fig:appendix:pattern-matching}
\end{figure}

\subsection{Modules and Definitions}
\label{appendix:semantics:mods-defs}
\begin{figure}
   \vspace{-1em}
   {\small \flushleft \shadebox{$\gamma, m, G \evalM v, G'$}%
      \begin{smathpar}
         \inferrule*[
            lab={\ruleName{$\evalM$-module}}
         ,  right={$n = \length{\blockSeq{d}}$}
         ]
         {
            \gamma_{i-1}, d_i \evalD \gamma_i, G_i
            \\
            \forall 1 \leq i \leq n
         }
         {
            \gamma_0, \blockSeq{d}, G_0 \evalM \gamma_n, G_n
         }
         \and
         \inferrule*[
            lab={\ruleName{$\evalM$-program}}
         ,  right={$n = \length{\blockSeq{d}}$}
         ]
         {
            \gamma_0, \blockSeq{d} \evalM \gamma_n, G_n
            \\
            s \desugar e
            \\
            \gamma_n, e, \varnothing, G_n \evalR v, G
            \\
            \forall 1 \leq i \leq n
         }
         {
            \gamma_0, \blockSeq{d}\;s \evalM v, G
         }
      \end{smathpar}
   }
   \\[-1cm]
   {\small \flushleft \shadebox{$\gamma, d, G \evalD \gamma\_, G\_$}%
      \begin{smathpar}
         \inferrule*[
            lab={\ruleName{$\evalD$-def}}
         ]
         {
            p \desugar x
            \\
            \gamma, e, V, G_1 \evalR v, G_2            
            \\
            }
            v, x \match \gamma\_, G_3
         {
            \gamma, \exDef{\sigma}{e}, V, G_1
            \evalD
            \gamma\_, G_2 \cup G_3
         }
         \and
         \inferrule*[
            lab={\ruleName{$\evalD$-fun-clause}}
         ]
         {
            \exApp{\gamma\_}{(x_i)} = \annClosure{\gamma}{\rho}{\exApp{\rho}{(x_i)}}{\alpha_i}
            \\
            \blockSeq{\clauseFun{x}{\seq{p}}{s}}, G \closeDefs \sigma, G\_
         }
         {
            \gamma, \blockSeq{\clauseFun{x}{\seq{p}}{s}} \evalD \gamma\_, G\_
         }
         \end{smathpar}
   }
   \caption{Evaluation of Modules and Programs}
   \label{fig:appendix:modules-defs}
\end{figure}

First we consider the evaluation relations for definitions $d$, as well as Modules ($\blockSeq{d}$), and programs ($\blockSeq{d}\;s$).
Fluid's module system is simple. When the compiler is applied to the program, it recursively evaluates the $\exImport{path}$ expressions
in the program, loading the mentioned modules in turn, and evaluating their own respective imports, constructing a graph. Once this
graph is produced, we attempt to perform a topological sort on the collection of modules. If this succeeds, we can proceed to construct
the environments in which we will run the final program, as well as the contextual graph from which we begin evaluation. One point of
interest is the fact that we do not transitively construct the environments. If $\kw{M}_a$ imports $\kw{M}_b$
and $\kw{M}_c$, the definitions in $\kw{M}_c$ are not visible from within $\kw{M}_a$, unless it imports $\kw{M}_c$ directly.
We shall omit the finer details of module importing since they are not essential for this work.
\clearpage

\section{Literate Execution Style in Python, Combining
Docstrings with Metaprogramming}
\label{app:literate-exe-python}

The following illustrates how to leverage Python's metaprogramming
facilities with docstrings to provide a simple literate execution
approach where docstrings containing variables are dynamically
evaluated and output to the user during program execution.
More sophisticated techniques, akin to those in Fluid, could
be developed from here; this is just a basic demonstration of
the idea.
\begin{python}
import inspect

def docformat(func):
    """Decorator to make docstrings format with function arguments."""
    def wrapper(*args, **kwargs):
        # Get argument names and values
        bound = inspect.signature(func).bind(*args, **kwargs)
        bound.apply_defaults()

        # Format docstring with available arguments
        if func.__doc__:
            formatted_doc = func.__doc__.format(**bound.arguments)
            print(formatted_doc)  # or store it somewhere
        return func(*args, **kwargs)
    wrapper.__doc__ = func.__doc__
    return wrapper

@docformat
def foo(x, y):
    """Foo called with {x} and {y}"""
    return x + y

print(foo.__doc__) # Prints "Foo called with {x} and {y}" 
foo(42, 100)       # Prints "Foo called with 42 and 100"
foo(1, 2)          # Prints "Foo called with 1 and 2"
\end{python}

\end{document}